\documentclass[twocolumn,notitlepage,nofootinbib,superscriptaddress]{revtex4-2}

\usepackage{amsmath,amssymb,bbm,soul}
\usepackage{amsthm}
\usepackage{xfrac}
\usepackage[colorlinks=true,citecolor=blue,linkcolor=blue,urlcolor=blue]{hyperref}

\catcode`\|=\active \def|{
\fontencoding{T1}\selectfont\symbol{124}\fontencoding{\encodingdefault}}
\newcommand{\nobracket}{}

\newcommand{\mathe}{\mathrm{e}}
\newcommand{\tmop}[1]{\ensuremath{\operatorname{#1}}}

\newcommand{\tmtextbf}[1]{\text{{\bfseries{#1}}}}

\usepackage{xcolor}

\usepackage[normalem]{ulem}
\newcommand{\stkout}[1]{\ifmmode\text{\sout{\ensuremath{#1}}}\else\sout{#1}\fi}

\begin{document}
\author{Nina Megier}

\email{nina.megier@mi.infn.it}

\affiliation{Dipartimento di Fisica “Aldo Pontremoli”, Università degli Studi di Milano, via Celoria 16, 20133 Milan, Italy}
\affiliation{ Istituto Nazionale di Fisica Nucleare, Sezione di Milano, via Celoria 16, 20133 Milan, Italy}
\affiliation{ International Centre for Theory of Quantum Technologies (ICTQT), University of Gdańsk, 80-308 Gdańsk, Poland}

\author{Andrea Smirne}
\affiliation{Dipartimento di Fisica “Aldo Pontremoli”, Università degli Studi di Milano, via Celoria 16, 20133 Milan, Italy}
\affiliation{ Istituto Nazionale di Fisica Nucleare, Sezione di Milano, via Celoria 16, 20133 Milan, Italy}

\author{Steve Campbell}

\affiliation{ School of Physics, University College Dublin, Belfield, Dublin 4, Ireland}
\affiliation{ Centre for Quantum Engineering, Science, and Technology, University College Dublin, Belfield, Dublin 4, Ireland}

\author{Bassano Vacchini}

\affiliation{Dipartimento di Fisica “Aldo Pontremoli”, Università degli Studi di Milano, via Celoria 16, 20133 Milan, Italy}
\affiliation{ Istituto Nazionale di Fisica Nucleare, Sezione di Milano, via Celoria 16, 20133 Milan, Italy}

\begin{abstract}
We critically examine the role that correlations established between a system and fragments of its environment play in characterising the ensuing dynamics. We employ a class of dephasing models where the state of the initial environment represents a tunable degree of freedom that qualitatively and quantitatively affects the correlation profiles, but nevertheless results in the same reduced dynamics for the system. We apply recently developed tools for the characterisation of non-Markovianity to carefully assess the role that correlations, as quantified by the (quantum) Jensen-Shannon divergence and relative entropy, as well as changes in the environmental state, play in whether the conditions for classical objectivity within the quantum Darwinism paradigm are met. We demonstrate that for precisely the same non-Markovian reduced dynamics of the system arising from different microscopic models, some will exhibit quantum Darwinistic features, while others show no meaningful notion of classical objectivity is present. Furthermore, our results highlight that the non-Markovian nature of an environment does not {\it a priori} prevent a system from redundantly proliferating relevant information, but rather it is the system's ability to establish the requisite correlations that is the crucial factor in the manifestation of classical objectivity.
\end{abstract}

\title{Correlations, information backflow, and objectivity in a class of pure dephasing models}

\maketitle

\section{Introduction}
The necessity for effective means to describe how a quantum system interacts with its surrounding environment has precipitated a burgeoning area of research. In many instances, one is solely focused on the dynamics of the system of interest, and therefore environmental effects can be phenomenologically modelled, rendering the complex system dynamics tractable~\cite{Breuer2002,Rivas2012}. While highly effective, such an approach neglects to account for the root of cause of the ensuing dynamics of the system. Reverting to a full microscopic description, where the system and environment interact and evolve according to an overall unitary dynamics, reveals that the correlations established between the system and the environment during their interaction play an important role in the resulting open dynamics of the system~\cite{Breuer2002,Rivas2012}. These correlations are the basis for notions of classical objectivity~\cite{Zurek2009a, HorodeckiPRA2015,  LePRA2018,LePRL2018, Korbicz2021a} and are also known to play a key role in the characterisation of the dynamics, in particular, if the system undergoes a Markovian (memoryless) or non-Markovian evolution~\cite{Rivas2014a,Breuer2016a}. {Both notions of classical objectivity and non-Markovian evolution have been the object of experimental investigations, 
see for example \cite{PhysRevA.98.020101,PhysRevLett.123.140402,CHEN2019580,Chisholm2022a} and \cite{Liu2011a,Rossi2017a,Liu2018,Cialdi2019a,White2020a,PhysRevA.104.022432,lyyra2021experimental} respectively.}

However, a given open system dynamics does not arise from a unique microscopic system-environment model, and rather there are infinitely many system-environment models that result the same system evolution~\cite{Smirne2021a}. Such an insight calls for a more careful analysis of the information exchanges between the system and its environment, allowing to more precisely pin down the relevant contributions which give rise to, e.g. Markovian vs. non-Markovian dynamics~\cite{Campbell2019b} or establish the conditions for classical objectivity~\cite{ZwolakNJP2012, KorbiczPRA19,SabrinaPRR}. This becomes particularly subtle since under such a microscopic picture the environment is typically composed of many constituent subsystems, and therefore it is relevant to assess the complementary role that global correlations established between the system and the whole environment play compared to correlations shared between the system and smaller environmental fragments. With regards to the former, it has recently been demonstrated that without the creation of strong global correlations in the form of entanglement, reasonable conditions for objectivity are not satisfied~\cite{SabrinaPRR, KorbiczPRA19}, while for the latter, it appears that only the correlations shared between the system and a small subset of the environmental degrees of freedom are relevant for the characterisation of the system dynamics~\cite{CampbellPRA2018, Campbell2019b}. 

In this work we attempt to unravel the contribution that various correlations play in the characterisation of an open system dynamics. To that end, we consider a spin-star dephasing model where several different initial environmental states, which in turn lead to significantly different correlation profiles, nevertheless produce the same reduced dynamics for the system~\cite{Smirne2021a}. We employ recently developed tools for understanding the emergence of non-Markovianity in terms of the correlations established between system and environment, as well as changes in the environmental state~\cite{Laine2010b, Campbell2019b}, to put into evidence the quite different role that these features play when characterising the dynamics, either in terms of its non-Markovian character or its ability to establish the conditions necessary for classical objectivity. We show that two different microscopic descriptions of the evolution that lead to the same reduced dynamics of the system can exhibit significant differences with regards to classical objectivity within the quantum Darwinism framework. Our work therefore demonstrates that the non-Markovian character of an evolution does not necessarily affect a system's ability to redundantly proliferate information to the environment, thus contributing to the ongoing efforts to unravel their relation~\cite{GiorgiPRA2015,Galve2016a,  GarrawayPRA2017, NadiaPRA} or possible lack thereof~\cite{LewensteinPRA2017, Ryan2021a}, note in particular the recent analysis in \cite{DiogoArXiv} complementary to ours.

The remainder of the paper is organised as follows. In Section~\ref{sec:class-deph-models} we introduce the class of spin-star models that will be our focus. Section~\ref{sec:corr} introduces the correlation measures that will be our key figures of merit and examines how they spread within different models. We analyse various information fluxes in Section~\ref{sec:nm} and their dual role characterising the non-Markovian nature of the dynamics and the redundant spreading of relevant system information to environmental constituents. Finally, we draw our conclusions in Section~\ref{sec:conc}.









\section{A class of dephasing models}\label{sec:class-deph-models}
Let us introduce the class of models for which we want to study the role of
correlations in determining important features of the overall and reduced
dynamics.
We recall that being interested in the reduced dynamics of the system in
a system-environment setting, the full specification of a model includes the
choice of the initial environmental state. We therefore consider a set of $N$
two-level systems with frequency $\omega_E$, interacting with a two-level system with frequency 
$\omega_S$, via the microscopic
Hamiltonian
\begin{align}
  H  =  \frac{\hbar \omega_S}{2} \sigma_z \otimes \mathbbm{1}_{2^N} &+
  \sum_{k = 1}^N g_k \sigma_z \otimes \sigma_z^k \nonumber \\&+ \sum_{k = 1}^N \frac{\hbar
  \omega_E}{2}  \mathbbm{1}_2 \otimes \sigma_z^k.  \label{eq:H}
\end{align}
With $\sigma_z^k$ we denote
the operator $\mathbbm{1}_{2^{(k - 1)}} \otimes \sigma_z \otimes
\mathbbm{1}_{2^{(N - k)}}$, where the Pauli matrix $\sigma_z$ is acting on the
$k$-th environmental qubit, while $\mathbbm{1}_d$ indicates the identity
operator in a space of dimension $d$, and the $g_k$'s are the system
environment coupling constants. Such an interaction corresponds to a
so-called spin-star setting, in which a central spin is coupled to
neighbouring environmental degrees of freedom which can be described by a
collection of non-interacting spins.
In particular, the considered coupling term is such that it only affects the
coherences of the system, since $\sigma_z \otimes \mathbbm{1}_{2^N}$ is a
constant of motion, thus describing a dephasing dynamics.

We will investigate the time evolution of these degrees of freedom in the
hypothesis of the existence of a closed reduced dynamics for the central spin
system, that is to say \ assuming the initial overall state factorized
according to $\rho_{SE} (0) = \rho_S (0) \otimes \rho_E (0)$. 
We will
consider a class of models in which the initial environmental state is given by a
tensor product of identical states, namely
\begin{eqnarray}
  \rho_E (0) & = & \bigotimes_{k = 1}^N \varrho_E,  \label{eq:rhoe}
\end{eqnarray}
with
\begin{eqnarray}
  \varrho_E & = &
                  \begin{pmatrix}
                             p & c\\
                             c & 1 - p
                 \end{pmatrix},
        \label{eq:ematrix}
\end{eqnarray}
where $p \in [0, 1]$ and without loss of generality we can take $c$ real in
the range $|c| \leqslant \sqrt{p (1 - p)}$. This class of initial
environmental states will allow us to explore not only the total correlations,
but also their establishment as a function of the fraction of environmental degrees of freedom we are taking into consideration.

Starting from the fact that the total unitary evolution operator in the
interaction picture can be written in the form
\begin{eqnarray}
  U ({s}) & = & \sum_{\{m_k \}} \mathe^{- i \sigma_z \left( \sum^N_{k = 1} g_k
  m_k \right) {s}} \otimes |\{m_k \} \rangle \langle \{m_k \}|,  \label{eq:u}
\end{eqnarray}
where the vectors $\{|\{m_k \} \rangle \nobracket = |m_1 \rangle \otimes
\ldots \nobracket \otimes |m_N \rangle \nobracket\}$, such that $\sigma_z^k
|m_k \rangle \nobracket = m_k |m_k \rangle \nobracket$ with $m_k \in \{- 1,
1\}$, denote the basis of eigenvectors of the operator $\bigotimes_{k = 1}^N
\sigma_z^k$ in the environmental space, we obtain for the evolved state of
system and environment the expression
\begin{eqnarray}
  \rho_{SE} ({s}) & = & 
                               \begin{pmatrix}
                                 \rho_{11} (0) \bigotimes_{k = 1}^N
                                 \rho_k ({s}) & \rho_{10} (0)
                                 \bigotimes_{k = 1}^N \sigma_k ({s})\\
                                 \rho_{01} (0) \bigotimes_{k = 1}^N
                                 \sigma_k^{\ast} ({s}) & \rho_{00} (0)
                                 \bigotimes_{k = 1}^N \rho_k^{\ast}
                                 ({s})
                               \end{pmatrix},
                                                       \label{eq:rest}
\end{eqnarray}
where
\begin{eqnarray}
  \rho_k ({s}) & = & 
                   \begin{pmatrix}
                     p & c \mathe^{- i 2 g_k {s}}\\
                     c \mathe^{i 2 g_k {s}} & 1 - p
                   \end{pmatrix}
  \label{eq:piu}
\end{eqnarray}
and
\begin{eqnarray}
 \sigma_k ({s}) & = &
  \begin{pmatrix}
      p \mathe^{- i 2 g_k {s}} & c\\
    c & (1 - p) \mathe^{i 2 g_k {s}}
  \end{pmatrix}.  \label{eq:meno}
\end{eqnarray}
An important feature of the considered class of evolutions appears when
considering the associated reduced dynamics. Indeed, taking the partial trace
with respect to the environmental degrees of freedom one immediately obtains
\begin{eqnarray}
  \rho_S ({s}) & = & 
                   \begin{pmatrix}
                     \rho_{11} (0) & \rho_{10} (0) \chi ({s})\\
                     \rho_{01} (0) \chi^{\ast} ({s}) & \rho_{00} (0)
                   \end{pmatrix}
\label{eq:rst}
\end{eqnarray}
with
\begin{eqnarray}
  \chi ({s}) & = & \prod_{k = 1}^N [\cos (2 g_k {s}) - i (2 p - 1) \sin (2 g_k
  {s})],  \label{eq:chi}
\end{eqnarray}
where we have used the identity
\begin{align}
  \sum_{\{ m_k \}} \mathe^{- i 2 \left( \sum^N_{k = 1} {g}_k m_k \right) {s}} \langle \{
m_k \} | \rho_{E } (0) | \{ m_k \} \rangle \nonumber\\=\prod_{k = 1}^N [\cos (2 {s}{g}_k) - i \langle \sigma_z^k \rangle_{\varrho_E} \sin (2 {s}{g}_k)],  \label{eq:cos}
\end{align}
and $\langle \ldots \rangle_{\varrho_E}$ denotes the expectation
value with respect to the state ${\varrho_E}$ given in Eq.~(\ref{eq:ematrix}),
so that the reduced dynamics is exactly the same for all initial environmental
states with the same diagonal matrix elements. Therefore, we have a whole class
of dephasing models, parametrized by the coherence, $c$, of the
environmental state given by Eq.~(\ref{eq:ematrix}), leading to exactly the
same reduced dynamics. The existence of different environments equally affecting a given system has been studied in different contexts, with the
purpose of allowing for more convenient 
numerical treatments
{\cite{Chin2010a,Martinazzo2011a,Tamascelli2018a,Tamascelli2019a}}. The
occurrence of the very same reduced evolution starting from different
microscopic dynamics in a controlled setting has been recently considered also
in {\cite{Smirne2021a}}, in order to investigate the physical mechanism behind
memory effects in a quantum dynamics.

For the sake of simplicity, and in order to allow for analytical results, we
will consider the case in which all coupling constants are taken to be equal
to a reference value $g$, so that all environmental units will evolve in the
same way throughout the dynamics, as well as a uniform distribution of the
populations in the initial environmental components, namely $p = 1 / 2$. In
particular, we will address situations in which $\varrho_E$ in
Eq.~(\ref{eq:ematrix}) ranges from pure, for $c=\pm\sfrac 12$, to maximally mixed
for $c = 0$. The maximally mixed state corresponds to a situation in which the
reduced environmental state is unaffected by the interaction with the
system.

We will now consider possible physical manifestations of the difference in the
microscopic dynamics and related correlations studying the onset of
Darwinistic behavior and non-Markovianity in the various environmental
scenarios.

\section{Spreading of correlations}\label{sec:corr}
Let us first study the establishment and spread of correlations in the
considered class of models. As discussed in detail in many publications~\cite{ZwolakNJP2012, Zwolak2013a, LePRL2018, SabrinaPRR, KorbiczPRA19, Cakmak2021a, CampbellPRA2019, MirkinEntropy, Touil2022a}, this
feature might have an impact on the notion of objectivity for a quantum state,
in the spirit of the so-called quantum Darwinism {\cite{Zurek2009a}} (see e.g.
{\cite{Korbicz2021a}} for a recent review and references therein). We will see
that it also provides us with interesting insights in the study of quantum
non-Markovianity {\cite{Rivas2014a,Breuer2016a}}.

\begin{figure*}
\hskip0.2\textwidth(\textbf{a})\hskip0.2\textwidth(\textbf{b})\hskip0.25\textwidth(\textbf{c})\\
\includegraphics[width=0.3\textwidth]{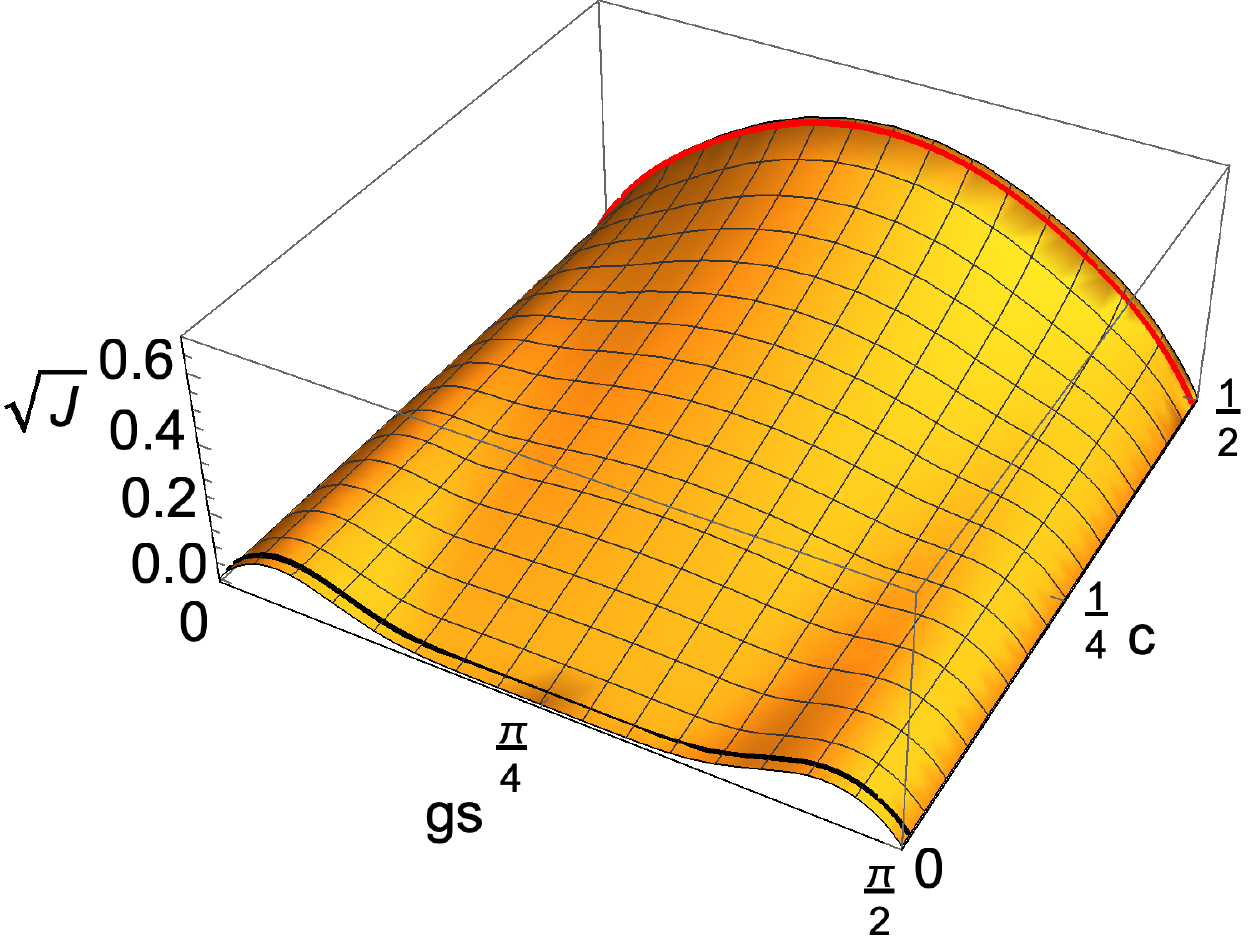}
\hspace{1em}
\includegraphics[width=0.3\textwidth]{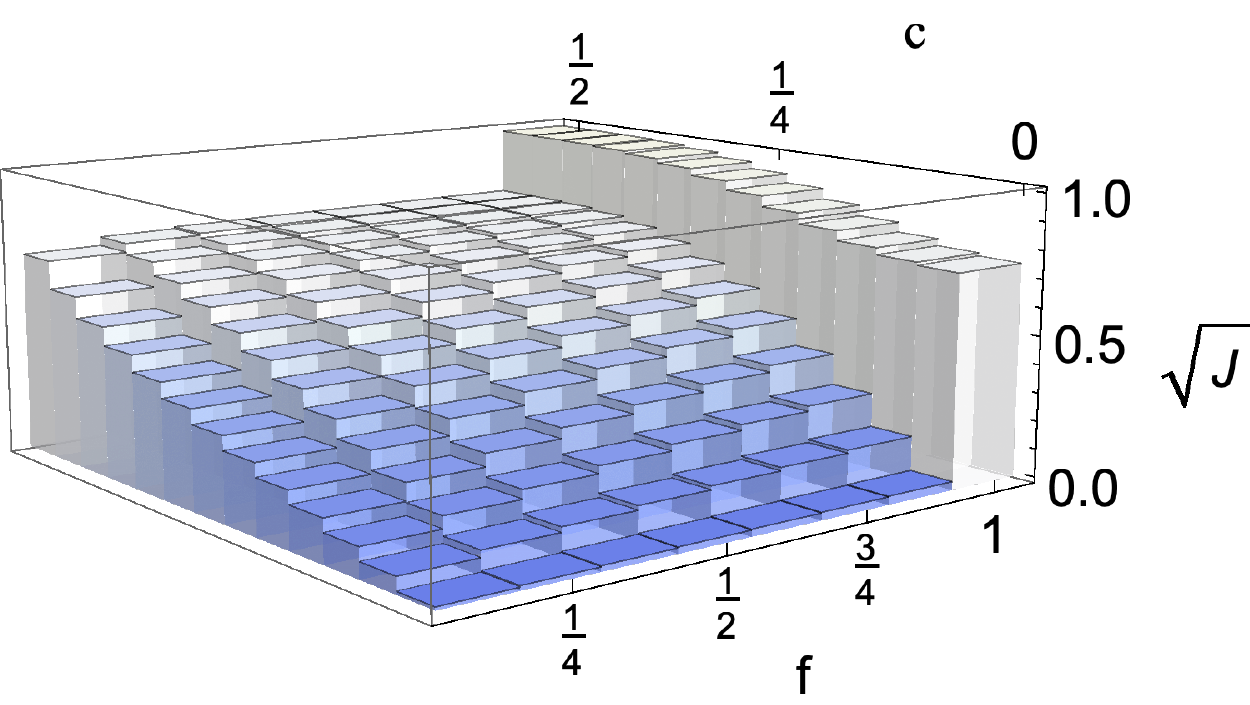}~~\includegraphics[width=0.3\textwidth]{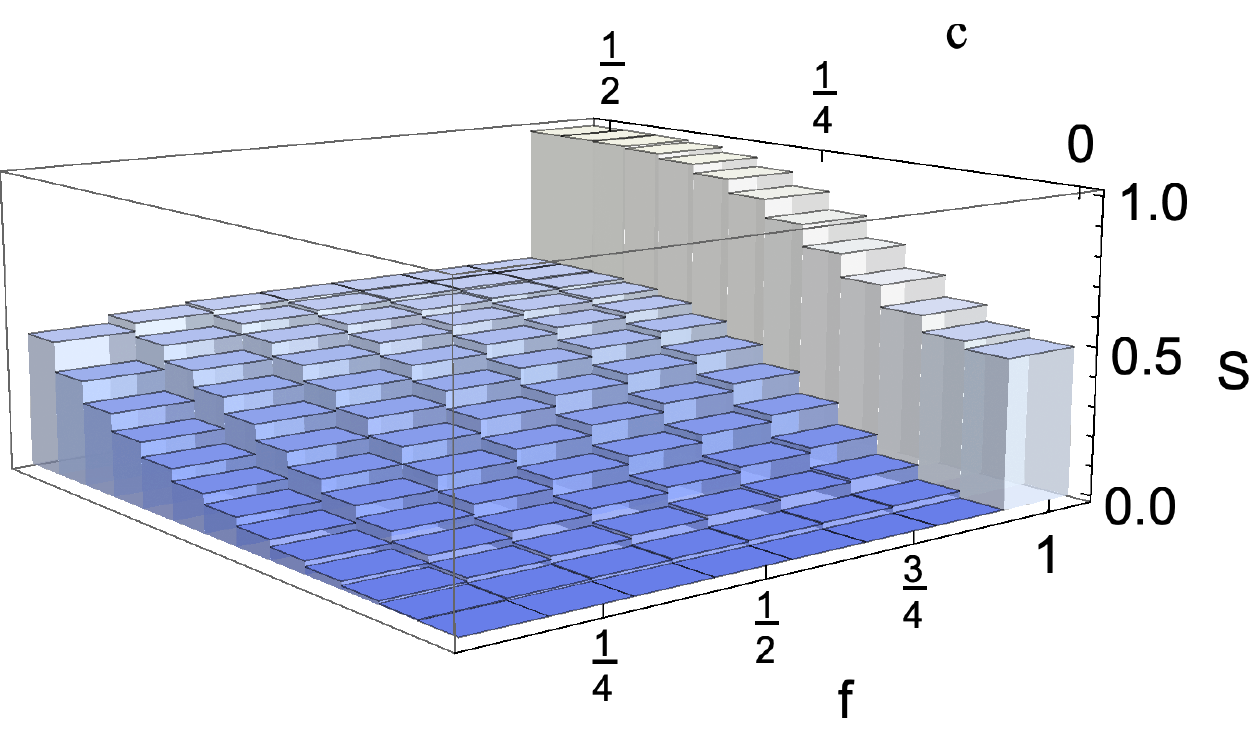}
\caption{(\textbf{a})  Amount of correlations shared between the
  system initially in the plus state $|+\rangle=(1/\sqrt{2} )(|0\rangle+|1\rangle)$  and one of the environmental qubits, evaluated by considering
  the QJSD$^{\sfrac 12}$ comparing this bipartite state with the
  product of its marginal as a function of time (in inverse units of
  the coupling parameter) and of the value $c$ of initial coherence in the
  environmental states. The quantity is renormalized to the value
  corresponding to a maximally entangled state.
Here and in the following figures the environment is composed of $N=8$
units. The black and red lines correspond to $c=0$ and $c=\sfrac 12$.
  (\textbf{b}) Distance between total state and product of its
  marginals at the reference time $gs=\pi/4$ as quantified by the
  QJSD$^{\sfrac 12}$, expressed as a function of the fraction of
  considered environmental qubits and of the value, $c$, of
  coherences in the environmental states. The total state includes the
  system and a fraction, ${\mathsf{f}}$, of the environmental qubits.
  (\textbf{c}) The same quantity obtained considering as quantifier
  the relative entropy, thus recovering the mutual information, still
  keeping the normalization to the value corresponding to the maximally
  entangled state. In both figures we see the emergence of a
  plateau for $c=\sfrac 12$, which is gradually washed out for smaller values
  of $c$, namely when moving from a model in which the environmental
  units have coherences to a fully diagonal state. 
}
\label{fig:plateaus}
\end{figure*}

As is clear from Eqs.~(\ref{eq:rst}) and (\ref{eq:chi}), for a uniform
coupling the reduced dynamics has a period of $\pi / 2$ in the
variable $g{s}$, so that we will consider times up to $\pi / (2 g)$. In
particular, the system is fully decohered for $g{s} = \pi / 4$. This decoherence
is connected to the establishment of correlations with the environmental
qubits, however, as shown in Fig.~\ref{fig:plateaus}, these correlations {(as quantified by the quantum Jensen-Shannon divergence defined in the following subsection)} are greater the more environmental qubits we take into account, in particular for $c=0$ the reduced system is only correlated at this point of time with the environment as a whole.
The overall state
according to Eq.~(\ref{eq:rest}) then reads
\begin{widetext}
\begin{equation}
  \rho_{SE} ({s}) = 
                      \frac{1}{2^N} 
                      \begin{pmatrix}
                        \rho_{11} (0)\bigotimes_{k = 1}^N
                        \begin{pmatrix}
                          1 & c \mathe^{- i 2 g  {s}}\\
                          c \mathe^{i 2 g{s}} & 1
                        \end{pmatrix}
                            & \rho_{10} (0)\bigotimes_{k = 1}^N 
                            \begin{pmatrix}
                              \mathe^{- i 2 g  {s}} & c\\
                              c & \mathe^{i 2 g{s}}
                            \end{pmatrix}
                            \\
                            \rho_{01} (0)\bigotimes_{k=1}^N
                                                                                                 \begin{pmatrix}
                                                                                                   \mathe^{i 2 g{s}} & c\\
                                                                                                   c
                                                                                                   &
                                                                                                   \mathe^{-
                                                                                                     i
                                                                                                     2
                                                                                                     g{s}}
                                                                                                   
                                                                                                 \end{pmatrix}
                                                                                                 & \rho_{00} (0)\bigotimes_{k = 1}^N 
                                                                                                 \begin{pmatrix}
                                                                                                   1 & c \mathe^{i 2 g{s}}\\
                                                                                                   c
                                                                                                   \mathe^{-
                                                                                                     i
                                                                                                     2
                                                                                                     g{s}}
                                                                                                   &
                                                                                                   1
                                                                                                 \end{pmatrix}
                                                                                                 \end{pmatrix}
\label{eq:c0t} .
\end{equation}
\end{widetext}
{Given that we are
considering a dephasing dynamics, a natural choice for the initial condition
for the system is a pure state of the form $\rho_S (0) = | + \rangle \langle +
|$, with $| + \rangle \nobracket = \left( 1 / \sqrt{2} \right) (|1 \rangle +
|0 \rangle)$, exhibiting the maximum amount of coherence, so that $\rho_{i j} (0)=\sfrac 12$ for $i,j=0,1$. 
Starting from this expression one can consider marginals in which less and
less environmental units are involved.} In particular, we will denote as
$\rho_{SE_{{\mathsf{f} N}}}$ the state obtained by tracing over all environmental
units not contained in a fraction $\mathsf{f}$ of the environment. For the
extreme cases $\mathsf{f} = 0$ and $\mathsf{f} = 1$ we recover the reduced and
total states, respectively.

\subsection{Quantifiers of correlations}
In order to understand the spreading of correlations in the different models
we are therefore interested in their dependence on the considered fraction. In
general, given a distinguishability quantifier between quantum states, say
$D$, that is a quantity defined on pairs of quantum states such that $D (\rho,
\sigma) \geqslant 0$, with equality iff the states coincide, one can use it as
a quantifier of correlations in a bipartite state considering the expression
$D (\rho_{SE}, \rho_S \otimes \rho_E)$. For the sake of this study, we will
consider the square root of the quantum Jensen-Shannon
divergence (QJSD$^{\sfrac 12}$) and the relative entropy. Both
will be used as quantifiers of bipartite correlations by renormalizing to the
value assumed for the case of a maximally entangled state. The choice of the
QJSD$^{\sfrac 12}$ is motivated by its use in the framework of
non-Markovianity {\cite{Megier2021a,Smirne2022a}}, while the relative entropy is typically
used in the framework of quantum Darwinism {\cite{Korbicz2021a}} due to its
connection with the mutual information. 

The QJSD$^{\sfrac 12}$ is defined in terms of the Jensen-Shannon divergence
{\cite{Bengtsson2017}} according to
\begin{eqnarray}
  \sqrt{J (\rho, \sigma)} & = & \sqrt{S \left( \frac{\rho + \sigma}{2} \right)
  - \frac{1}{2} S (\rho) - \frac{1}{2} S (\sigma)},  \label{eq:j}
\end{eqnarray}
where $S (\rho) = - \tmop{Tr} \rho \log \rho$ denotes the von Neumann entropy
and logarithms are considered in base 2. This quantity, besides being a
well-known distinguishability quantifier, has recently been shown to be a
distance {\cite{Sra2021a,Virosztek2021a}}. In particular when used to evaluate
correlations it takes the form
\begin{align}
   \sqrt{J (\rho_{SE}, \rho_S \otimes \rho_E)} =\nonumber \\
   \sqrt{S \left(
  \frac{\rho_{SE} + \rho_S \otimes \rho_E}{2} \right) - \frac{1}{2} S
(\rho_{SE}) - \frac{1}{2} S (\rho_S) - \frac{1}{2} S (\rho_E)},
\nonumber
\end{align}
taking the value $\sqrt{2 - (5 / 8) \log 5} \approx 0.74$ when $\rho_{SE}$
corresponds to the maximally entangled state in $\mathbbm{C }^2 \otimes
\mathbbm{C}^{2 N}$. We will denote as $\sqrt{\mathsf{J}}$ the quantity
rescaled by this factor, thus assuming unity for maximally entangled
states.

The relative entropy is defined according to {\cite{Bengtsson2017}}
\begin{eqnarray}
  S (\rho, \sigma) & = & \tmop{Tr} \rho \log \rho - \tmop{Tr} \rho \log
  \sigma, \label{eq:s} 
\end{eqnarray}
so that when used to quantify correlations it leads to the mutual information
\begin{eqnarray}
  S (\rho_{SE}, \rho_S \otimes \rho_E) & = & S (\rho_S) + S (\rho_E) - S
  (\rho_{SE}) \label{eq:mi}, 
\end{eqnarray}
providing a natural quantifier of both classical and quantum
correlations.
Considering again logarithms in base 2 it takes the value 2 for the maximally entangled state in $\mathbbm{C }^2
\otimes \mathbbm{C}^{2 N}$, so that we will denote as $\mathsf{S}$ the
quantity rescaled by a factor 2.

\subsection{Model dependence of correlation formation}
The key quantities to be considered in the study of the establishment of
correlations between the system and different parts of the environment in the
different considered models are therefore $\sqrt{\mathsf{J}
(\rho_{SE_{{\mathsf{f} N}}}, \rho_S \otimes \rho_{E_{{\mathsf{f} N}}})}$ and
$\mathsf{S} (\rho_{SE_{{\mathsf{f} N}}}, \rho_S \otimes \rho_{E_{{\mathsf{f} N}}})$.
Their behavior is shown in Fig.~\ref{fig:plateaus}(\tmtextbf{b}) and
Fig.~\ref{fig:plateaus}(\tmtextbf{c}) respectively, as a function of the
parameter $c$, which fixes the initial environmental state and therefore the
model. As follows from their expressions given in Eqs.~(\ref{eq:j}) and
(\ref{eq:mi}), their determination relies on knowledge of the eigenvalues of
$\rho_{SE_{{\mathsf{f} N}}}$, $\rho_S \otimes \rho_{E_{{\mathsf{f} N}}}$ and their
average. In turn, these operators depend on the chosen initial state for the
system, that we have taken to be the pure state $\rho_S (0) = | + \rangle
\langle + |$, initially exhibiting the maximum amount of coherence, so as to better put into evidence the role of the environment.

For the case $c = 0$ one immediately sees from Eq.~(\ref{eq:c0t}) that the
environment is left unchanged, so that it remains in the maximally mixed
state. The eigenvalues of the relevant states can be shown to be
\begin{align}
  &\rho_{SE_{{\mathsf{f} N}}} ({s})  \rightarrow  \frac{1}{2^{{{\mathsf{f} N}} + 1}} (1 \pm
  \cos^{N - {{\mathsf{f} N}}} (2 g{s})) \label{eq:eigen0} \\
  &\rho_S ({s})  \rightarrow  \frac{1}{2} (1 \pm \cos^N (2 g{s})) \nonumber\\
  &\rho_{E_{{\mathsf{f} N}}} ({s})  \rightarrow  \frac{1}{2^{{\mathsf{f} N}}} \nonumber\\
  &\frac{\rho_{SE_{{\mathsf{f} N}}} ({s}) + \rho_S({s}) \otimes
  \rho_{E_{{\mathsf{f} N}}}({s})}{2}  \rightarrow 
	\nonumber \\ 
	&\frac{1}{2^{{{\mathsf{f} N}} + 1}} \left( 1 \pm
  \frac{1}{2} \left \lvert  \cos^N (2 g{s}) + \cos^{{(1-{\mathsf{f}) N}}} (2 g{s}) \mathe^{i 2 g\left(
  \sum_{k = 1}^{{\mathsf{f} N}} m_k \right) s} \right  \rvert \right) \nonumber
\end{align}
where the eigenvalues for $\rho_{SE_{{\mathsf{f} N}}} ({s})$ and
$\rho_{E_{{\mathsf{f} N}}} ({s})$ are $2^{{\mathsf{f} N}}$ degenerate, while the numbers $\{ m_k
\}$ belong to $\{- 1, 1\}$ and their value is determined by the associated
eigenvector. In terms of these expressions, exploiting the fact that the von
Neumann entropy of a state only depends on its eigenvalues,
\begin{eqnarray}
  S (\rho) & = & - \sum_i \rho_i \log \rho_i, 
\end{eqnarray}
one can analytically determine the relevant expressions for the correlations.

An arbitrary value of the coherences in the initial environmental state calls
for a numerical evaluation, whose results are shown in
Fig.~\ref{fig:plateaus}(\tmtextbf{b}) and
Fig.~\ref{fig:plateaus}(\tmtextbf{c}) at time $g{s} = \pi / 4$, when the system
has fully decohered, as can be seen from Eqs.~(\ref{eq:rst}) and
(\ref{eq:chi}), thus loosing its initial information content. It
immediately appears, independently of the chosen correlation quantifier, that
for $c = \sfrac 12$, i.e. initially pure environmental units, the system shares an
equal amount of correlations with any small fraction of the environment,
corresponding to a plateau in the fraction dependence of the correlation
quantifiers. This feature is interpreted in the literature as quantum
Darwinism {\cite{Zurek2009a}}, namely a redundant storing of information about
the system in different portions of the environment, allowing for a notion of
objectivity, in the sense that the same information can be retrieved by
different observers accessing distinct parts of the environment. It is to be
stressed that the mutual information provides the standard choice of correlation
quantifier in this framework, though others have also been considered
{\cite{Zwolak2013a,Touil2022a}}. This notion of objectivity is not
uncontroversial, see {\cite{Korbicz2021a}} for a critical discussion and
further developments. The formation of the plateau is slowed down with
decreasing $c$, while for $c = 0$, such that the environmental units are
maximally mixed, correlations are only established between the system and the
environment as a whole. This behavior, namely the gradual washing out of
Darwinism in the dependence on the state of the environmental units was
already observed in {\cite{Zwolak2009a}}, taking as a figure of merit to
characterize the effect the von Neumann entropy of the units, which is a
monotonic function of their coherences, see also~\cite{, ZwolakPRA2010, KorbiczPRL2014}. 

To exemplify the distribution of
correlations for the case $c = 0$, let us write the overall state
Eq.~(\ref{eq:c0t}) for the case of two environmental qubits, thus
obtaining
\begin{widetext}
\begin{align}
  \rho_{SE} ({s}) = \frac{1}{8} \begin{pmatrix}
    1 & e^{- i 4 g{s}}\\
    e^{i 4 g{s}} & 1
  \end{pmatrix} \otimes \begin{pmatrix}
    1 & 0 & 0 & 0\\
    0 & 0 & 0 & 0\\
    0 & 0 & 0 & 0\\
    0 & 0 & 0 & 0
  \end{pmatrix}   + \frac{1}{8} \begin{pmatrix}
    1 & 1\\
    1 & 1
  \end{pmatrix} \otimes \begin{pmatrix}
    0 & 0 & 0 & 0\\
    0 & 1 & 0 & 0\\
    0 & 0 & 1 & 0\\
    0 & 0 & 0 & 0
  \end{pmatrix}\nonumber + \frac{1}{8} \begin{pmatrix}
    1 & e^{i 4 g{s}}\\
    e^{- i 4 g{s}} & 1
  \end{pmatrix} \otimes \begin{pmatrix}
    0 & 0 & 0 & 0\\
    0 & 0 & 0 & 0\\
    0 & 0 & 0 & 0\\
    0 & 0 & 0 & 1
  \end{pmatrix}, 
\end{align}
\end{widetext}
namely a classically correlated state, apart from multiples of $g{s} = \pi / 2$,
when the factorized initial state is recovered due to periodicity. Tracing out
all but one
of the environmental qubits, we obtain
\begin{align}
  \rho_{SE_{1}} ({s}) &= \frac{1}{8} \begin{pmatrix}
    2 & 1 + e^{- i 4 g{s}}\\
    1 + e^{i 4 g{s}} & 2
  \end{pmatrix} \otimes \begin{pmatrix}
    1 & 0\\
    0 & 0
  \end{pmatrix}\nonumber \\
& + \frac{1}{8} \begin{pmatrix}
    2 & 1 + e^{i 4 g{s}}\\
    1 + e^{- i 4 g{s}} & 2
  \end{pmatrix} \otimes \begin{pmatrix}
    0 & 0\\
    0 & 1
  \end{pmatrix}, 
\end{align} 
which is immediately seen to be factorized for $g s = \pi / 4$, i.e. when the
system has fully decohered. Accordingly, the information about the reduced
system is stored then solely in the global correlations between the
system and the environment, all partial fractions of the environment are not
correlated with the reduced system (note that the reduced density matrix of the
environment is always maximally mixed). This is exactly the behavior appearing in
Fig.~\ref{fig:plateaus}(\tmtextbf{b}) and
Fig.~\ref{fig:plateaus}(\tmtextbf{c}).


\begin{figure}
\begin{center}
    \includegraphics[width=0.75\columnwidth]{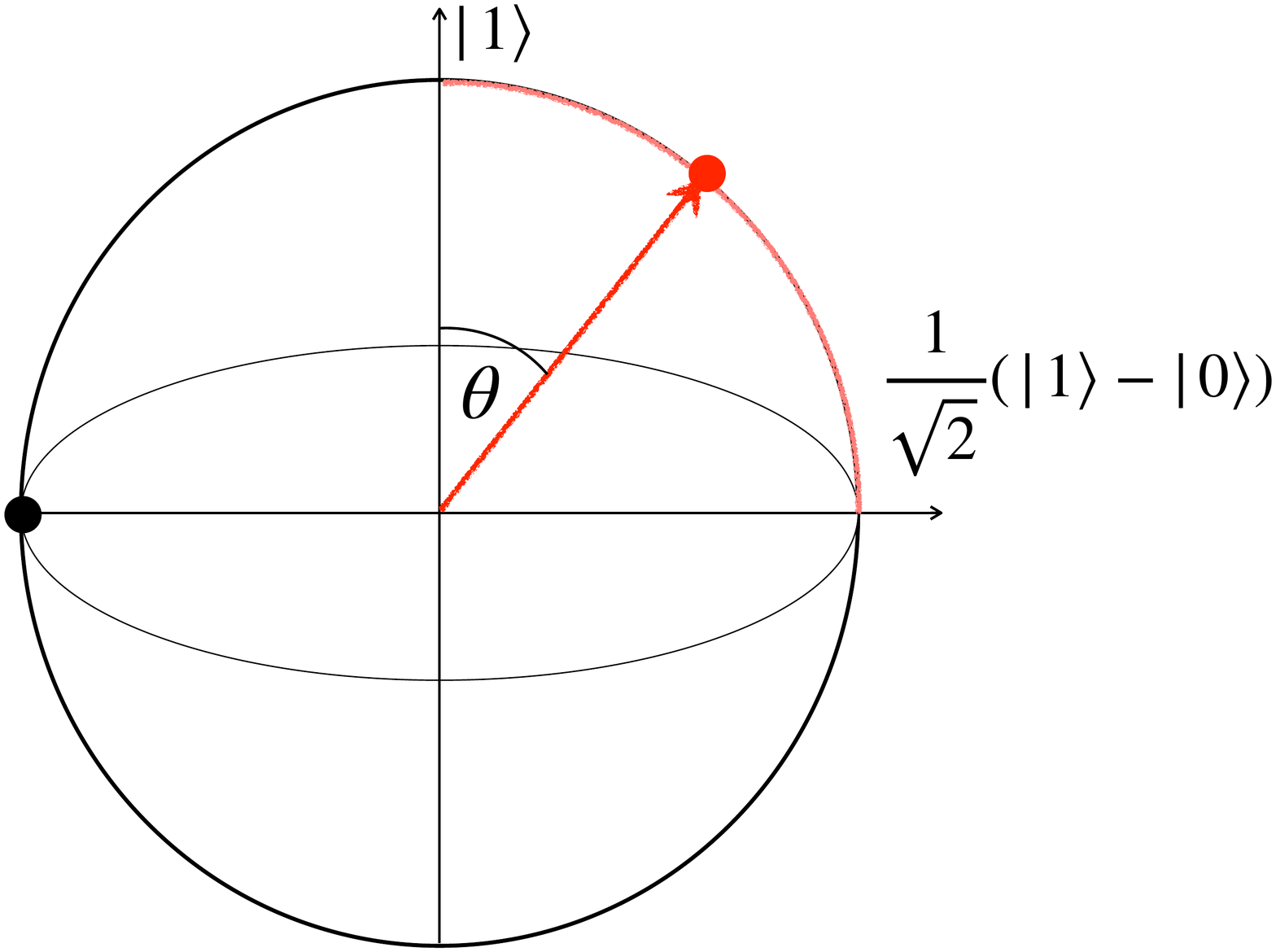}
\end{center}
    \caption{Bloch sphere representation of the considered pair of initial system states. One state is fixed to be the equatorial plus state $|+\rangle=(1/\sqrt{2} )(|0\rangle+|1\rangle)$ (black dot), while the other element of the pair belongs to the maximum circle and is characterized by the angle $\theta$ (red dot). For $\theta=\pi/2$  it becomes the minus state $|-\rangle=(1/\sqrt{2} )(|0\rangle-|1\rangle)$ and one recovers an orthogonal pair of initial states. For $\theta=0$ it corresponds to the up state $|1\rangle$.}
\label{fig:bloch}
\end{figure}

\section{Information backflow}\label{sec:nm}
We now want to analyze the features of the different models in the framework
of non-Markovianity, that is the study of memory effects in a quantum setting.
In this respect we will make reference to an approach to quantum
non-Markovianity focusing on features of the reduced dynamics
{\cite{Rivas2014a,Breuer2016a}}, at variance with viewpoints which more
closely mimic the classical definition of a non-Markovian process referring
to joint probability distributions {\cite{Pollock2018a}}, thus involving
information on intermediate steps necessary in order to extract information
from a quantum system {\cite{Vacchini2011a}}. Given that the considered
definition of non-Markovian dynamics will only involve the reduced dynamics,
the whole class of considered models will perform in the exactly same way.
Nevertheless the definition to be considered relies on the information
exchange between system and environment, which manifests differently in the
various models.

Let us first briefly formalize the pioneering approach to non-Markovianity of
a quantum dynamics introduced in {\cite{Breuer2009b,Laine2010a}}. The basic idea is to
consider the evolution in time of the distinguishability between two system
states, associating to a non-monotonicity in time of this quantity the
definition of non-Markovian dynamics. The motivation behind this definition is
that revivals of distinguishability can be unambiguously associated to
information backflow from external degrees of freedom to the system. This
approach has been initially formulated in terms of the trace distance
{\cite{Breuer2016a}}, but is actually amenable to the use of other
distinguishability quantifiers, in particular entropic ones, which come closer
to the present treatment focused on the spreading of correlations, as shown in
{\cite{Megier2021a,Smirne2022a}}.
\begin{figure}
\begin{center}
             \includegraphics[width=0.75\columnwidth]{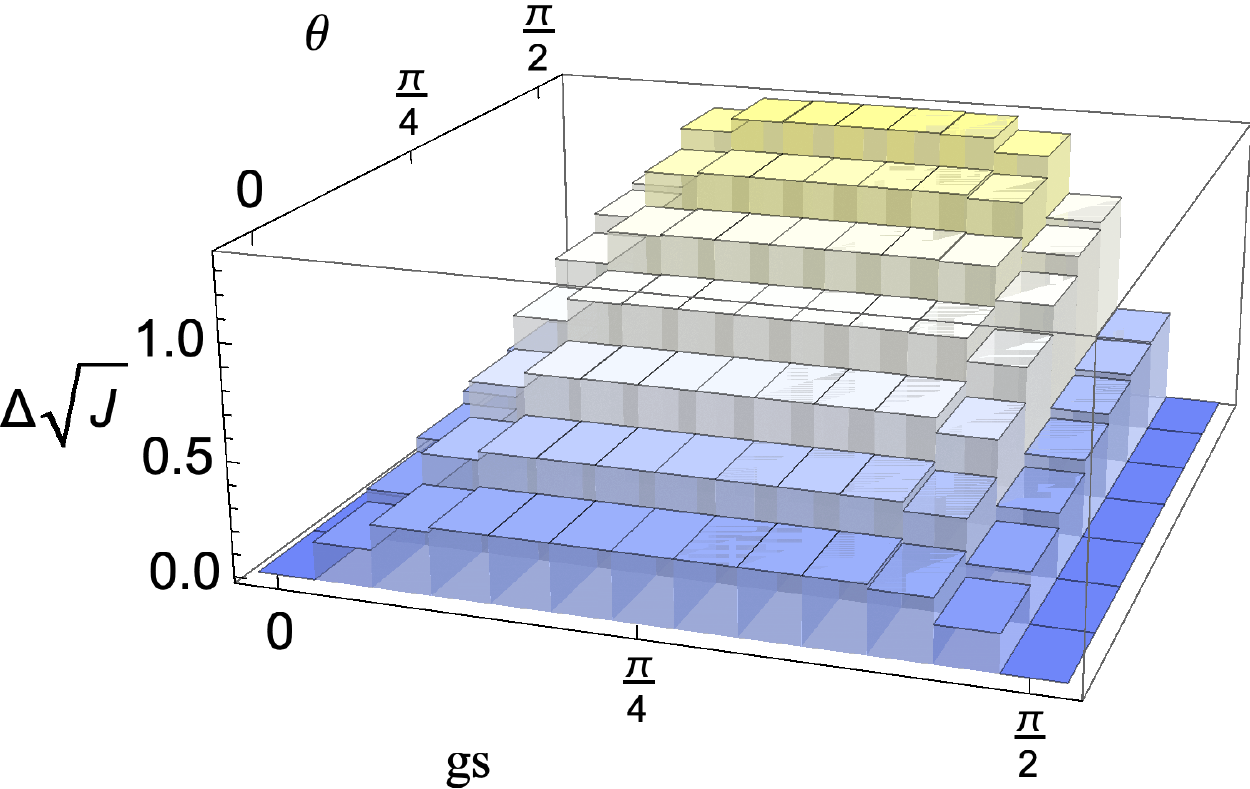}       
\end{center}
             \caption{Plot of the l.h.s. of Eq.~(\ref{eq:ineq})
               showing the revivals of the QJSD$^{\sfrac 12}$ as a
               function of time and choice of initial system states. The
               reference time $gt$ is fixed to be $\pi/2$,  i.e. after one full period of the evolution, while $gs$
               sweeps from 0 to $\pi/2$. The initial pair of system
               states is given by $\rho^1_S (0) = | + \rangle \langle
               + |$ and $\rho^2_S (0) =| \theta \rangle \langle \theta
               |$, as shown in Fig.~\ref{fig:bloch}, with $\theta$ ranging from 0 to $\pi/2$,
               corresponding to the case of an orthogonal pair and
               maximizing the revivals.}
  \label{fig:lhs_time}
\end{figure}
The key quantity to be considered is therefore the distinguishability of two
system states evolved from two distinct initial conditions, namely
\begin{equation}
  \sqrt{\mathsf{J} (\rho^1_S ({s}), \rho^2_S ({s}))}, \label{eq:js1s2}
\end{equation}
where as discussed we have used as distinguishability quantifier the
QJSD$^{\sfrac 12}$ as defined in Eq.~(\ref{eq:j}). The QJSD$^{\sfrac 12}$
is a contraction with respect to the action of any positive trace
preserving map, so that $\sqrt{\mathsf{J} (\rho^1_S ({s}), \rho^2_S ({s}))}$ is
monotonically decreasing in the case of a 
positive divisible evolution. For
more general dynamics, this quantity can show revivals in time, pointing to
the existence of memory effects. 
{In particular the revivals from the value at a time $s$ to a value at a later time $t$
can be upper bounded according to
\begin{align}\label{eq:ineq}
\sqrt{\mathsf{J} (\rho^1_S (t), \rho^2_S (t))} - \sqrt{\mathsf{J}
  (\rho^1_S (s), \rho^2_S (s))} \leqslant  \sqrt{\mathsf{J} (\rho^1_E (s),
  \rho^2_E (s))} 
	\nonumber  \\ 
  + \sqrt{\mathsf{J} (\rho^1_{SE} (s), \rho^1_S (s) \otimes \rho^1_E (s))}
  + \sqrt{\mathsf{J} (\rho^2_{SE} (s), \rho^2_S (s) \otimes \rho^2_E (s))},
\end{align}
where $\rho^{1, 2}_E ({s})$ denote the time evolved
environmental states corresponding to the initial condition $\rho^{1, 2}_S
(0)$, while {$\rho^{1}_E (0)=\rho^{2}_E (0)$ is} determined as above by fixing the model of
interest.} Given that all three contributions at the r.h.s. are zero if and
only if their arguments are equal, this bound has a clear physical meaning:
Non-Markovianity as described by revivals in the distinguishability of system
states can only take place if correlations have been established
between system and environment which is captured by the last two terms on the r.h.s of Eq.~\eqref{eq:ineq} and/or different initial system states have affected, in a
different way, the state of the environment, captured by the first term on the r.h.s. of Eq.~\eqref{eq:ineq}. In both cases some information has
been stored in degrees of freedom external with respect to the system. The
revivals do depend, in general, on the choice of initial system states, so that
it is natural to consider initial pairs that can be perfectly distinguished,
namely orthogonal states. In our case, given the previously considered choice
$\rho^1_S (0) = | + \rangle \langle + |$, this would amount to considering
$\rho^2_S (0) = | - \rangle \langle - |$. However, one immediately realizes
that in analogy to the fact that the reduced system dynamics is only affected
by the diagonal matrix elements of the environmental qubits, also the dynamics
of the environmental states only depends on the diagonal elements of
$\rho^{1, 2}_S (0)$ in the $\sigma_z$ basis. This would automatically imply
the vanishing of the first term at the r.h.s. of Eq.~(\ref{eq:ineq}). We will
therefore consider a more general pair of initial states, namely $\rho^1_S
(0) = | + \rangle \langle + |$ and $\rho^2_S (0) = | \theta \rangle \langle
\theta |$, with $| \theta \rangle \nobracket = \cos (\theta / 2) |1 \rangle -
\sin (\theta / 2) |0 \rangle$, as depicted in Fig.~\ref{fig:bloch}. 

We now
want to explore the behavior of these bounds for the different considered
microscopic models, investigating in particular what happens when only partial
information on the environment can be obtained.
\begin{figure*}
  \begin{flushleft}
    $\mathbf{c=\sfrac 12}$
  \end{flushleft}
 \includegraphics[width=0.24\textwidth]{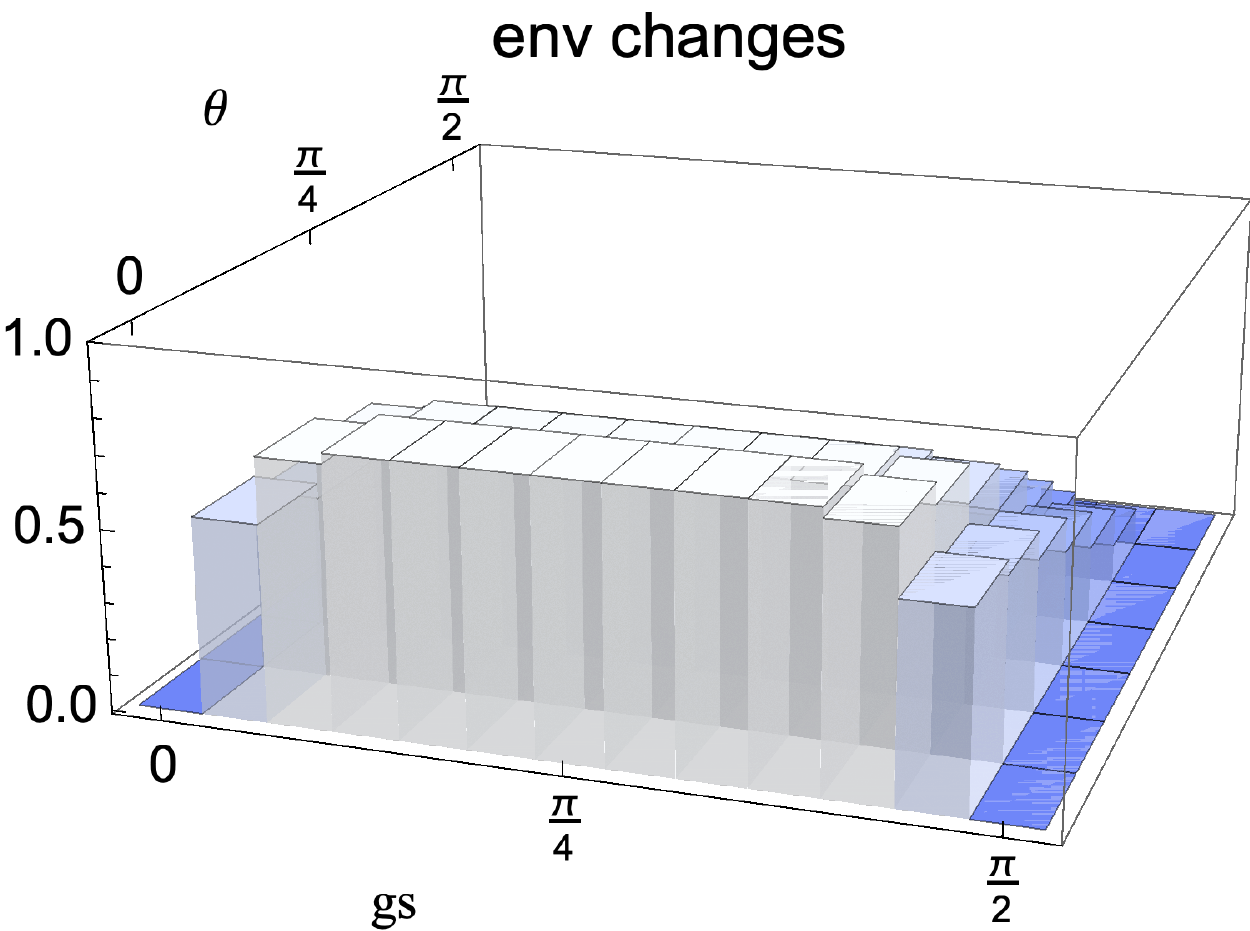}~~\includegraphics[width=0.24\textwidth]{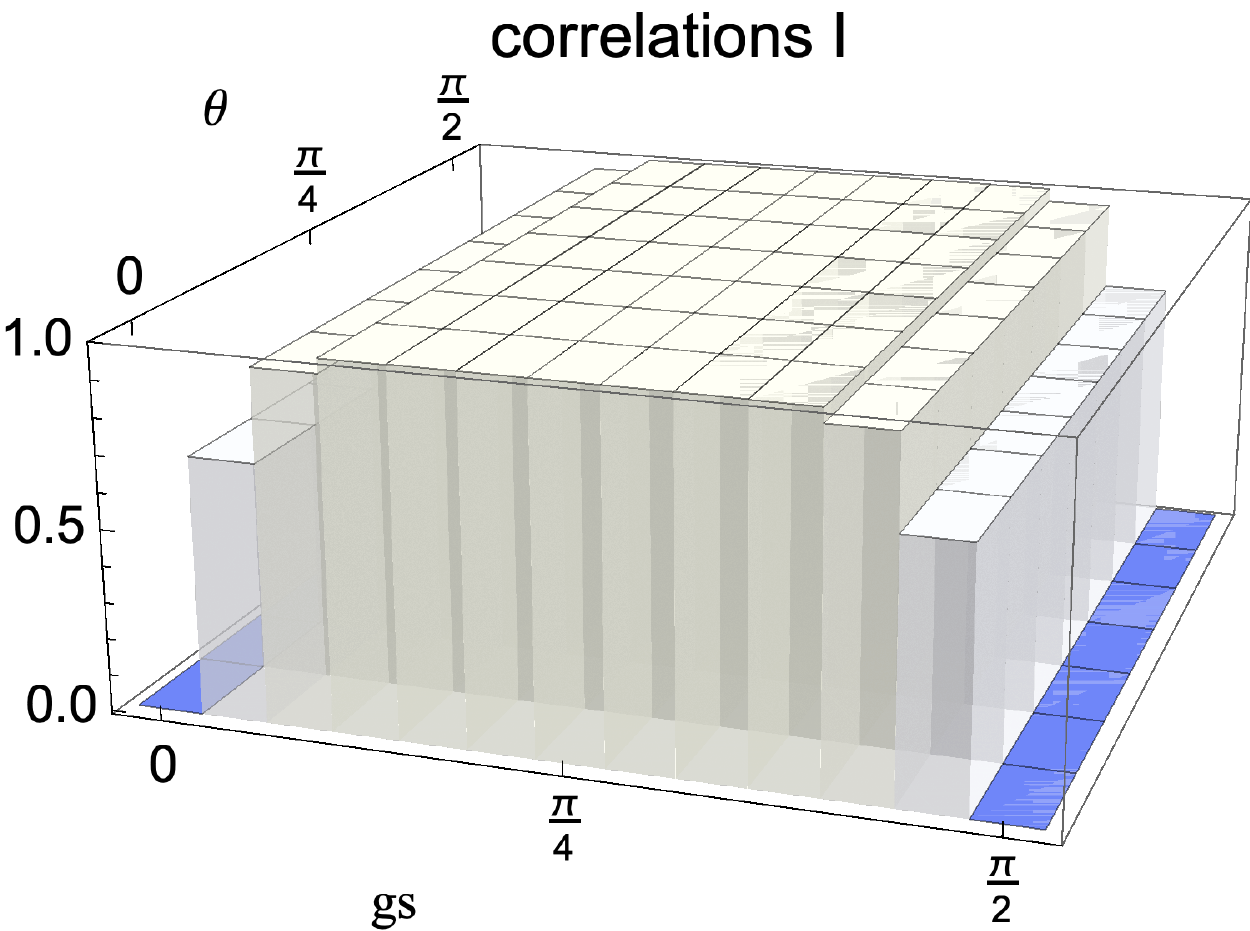}~~\includegraphics[width=0.24\textwidth]{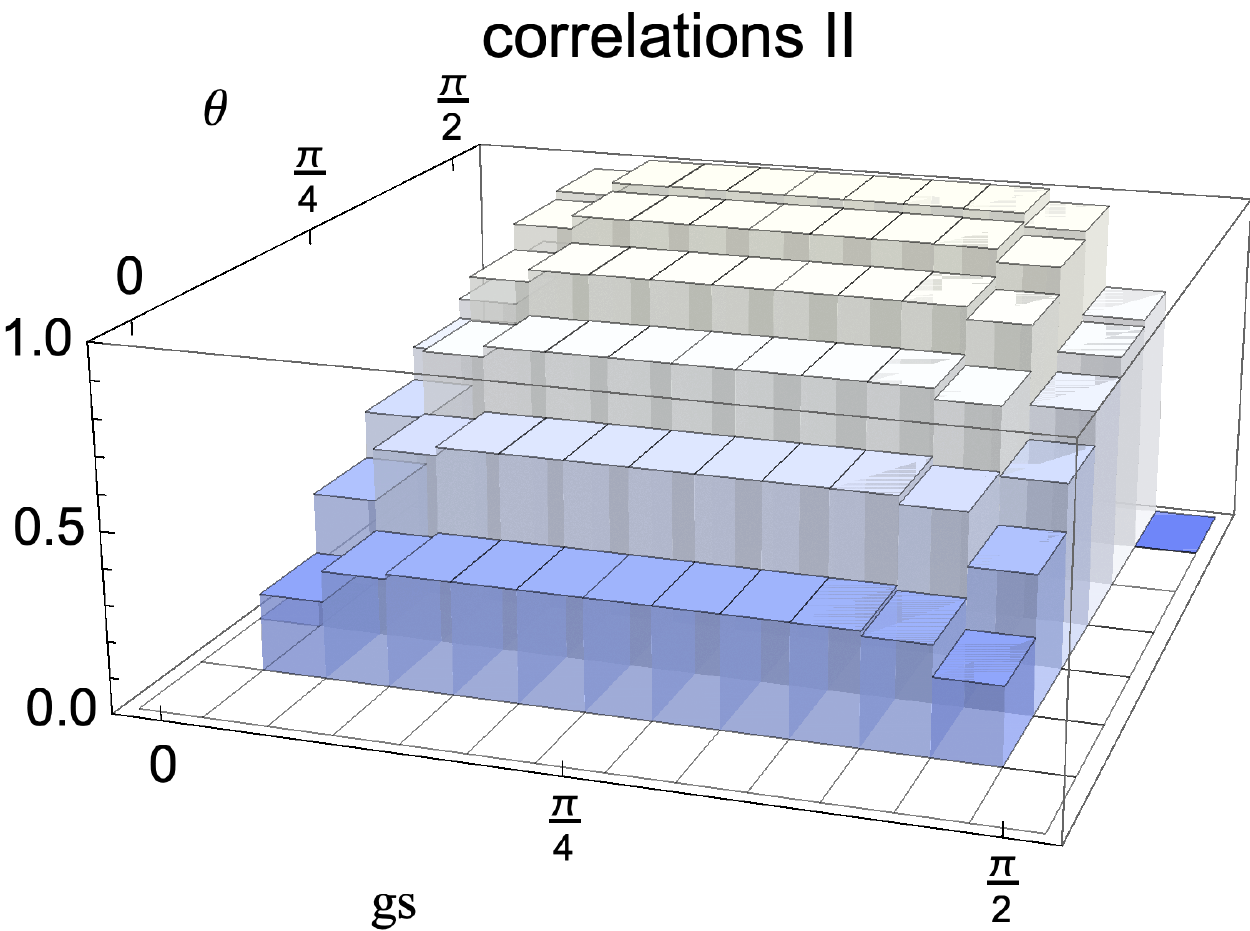}~~\includegraphics[width=0.24\textwidth]{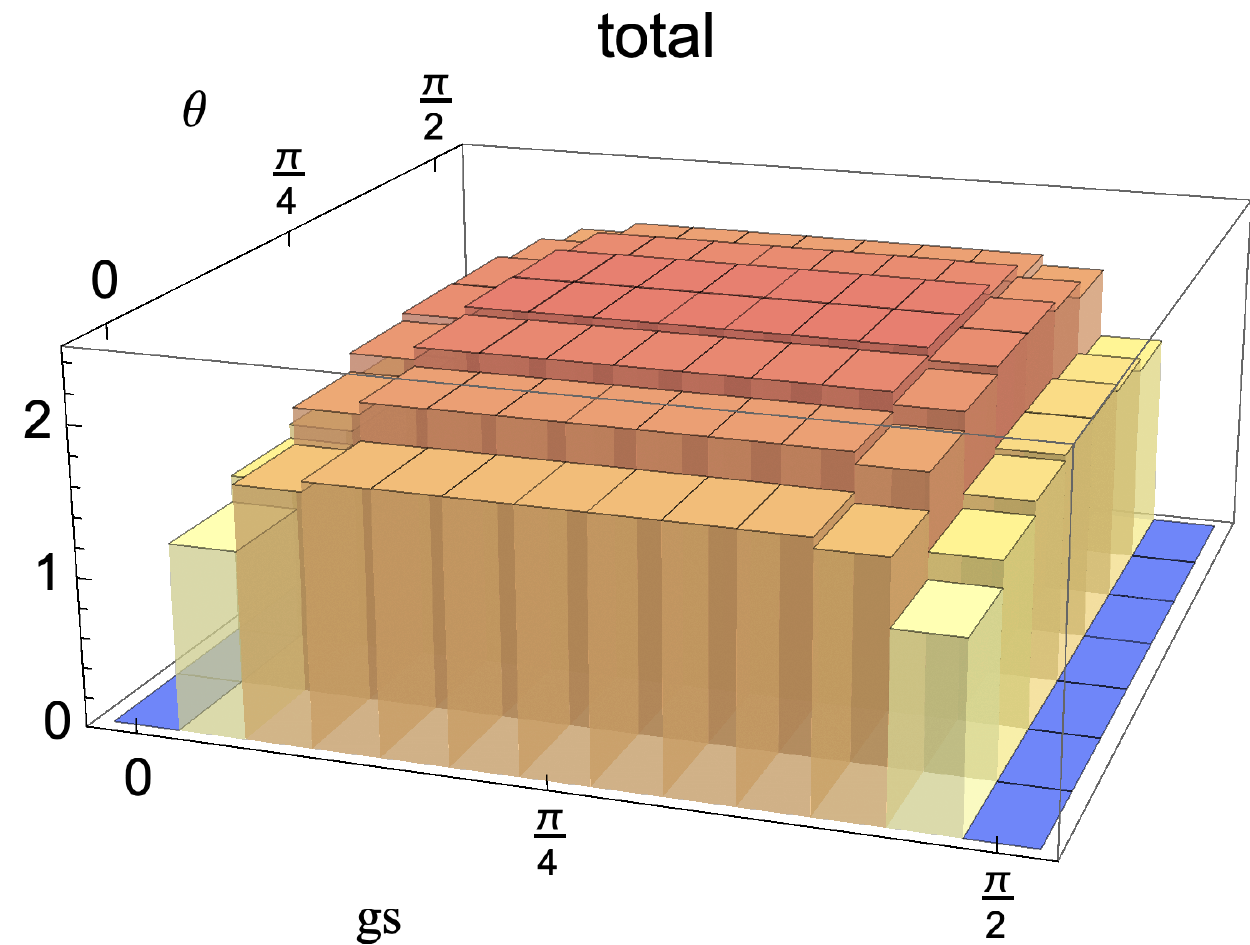}
\rule{\textwidth}{.1em}
\begin{flushleft}
    $\mathbf{c=0}$
  \end{flushleft}
  \includegraphics[width=0.24\textwidth]{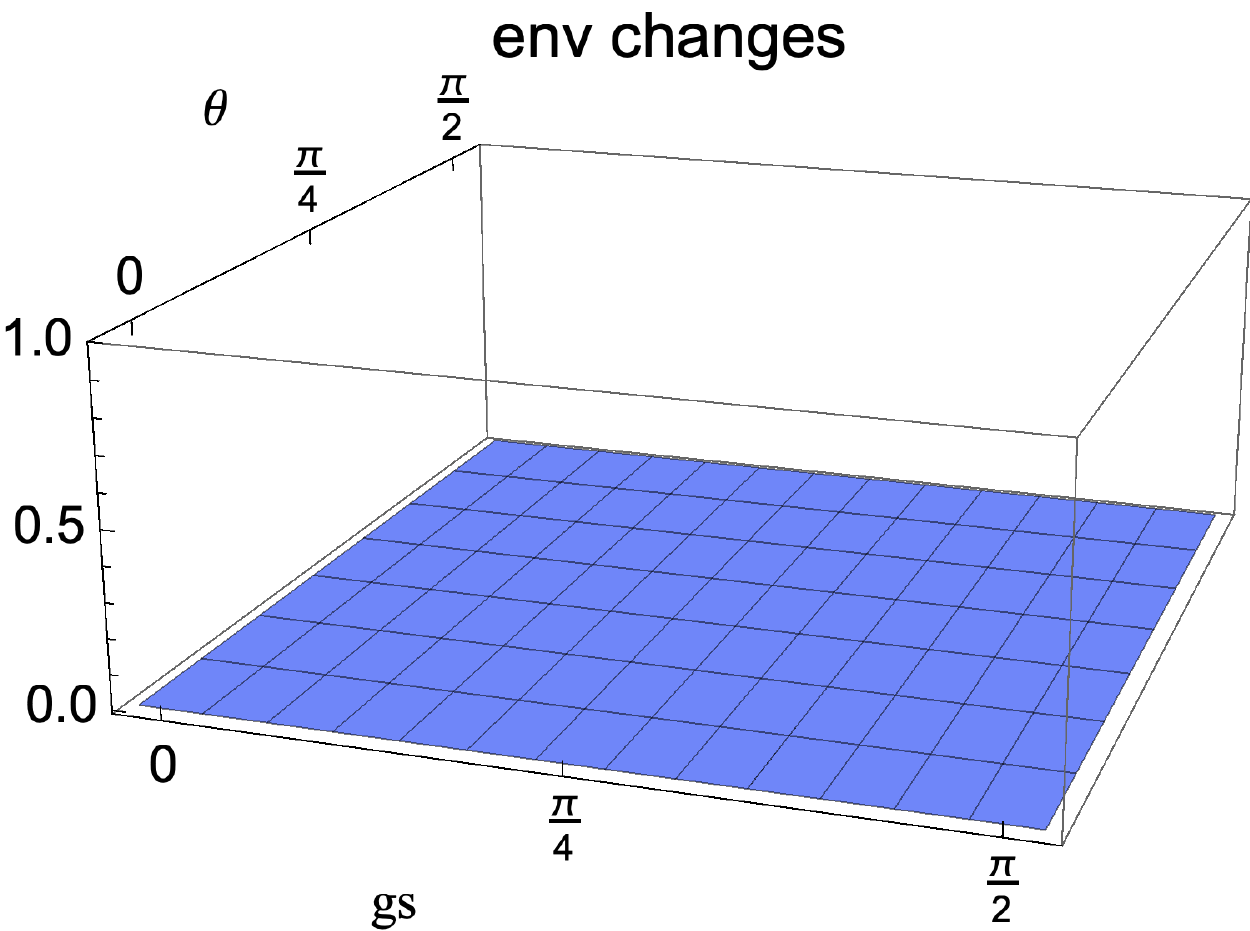}~~\includegraphics[width=0.24\textwidth]{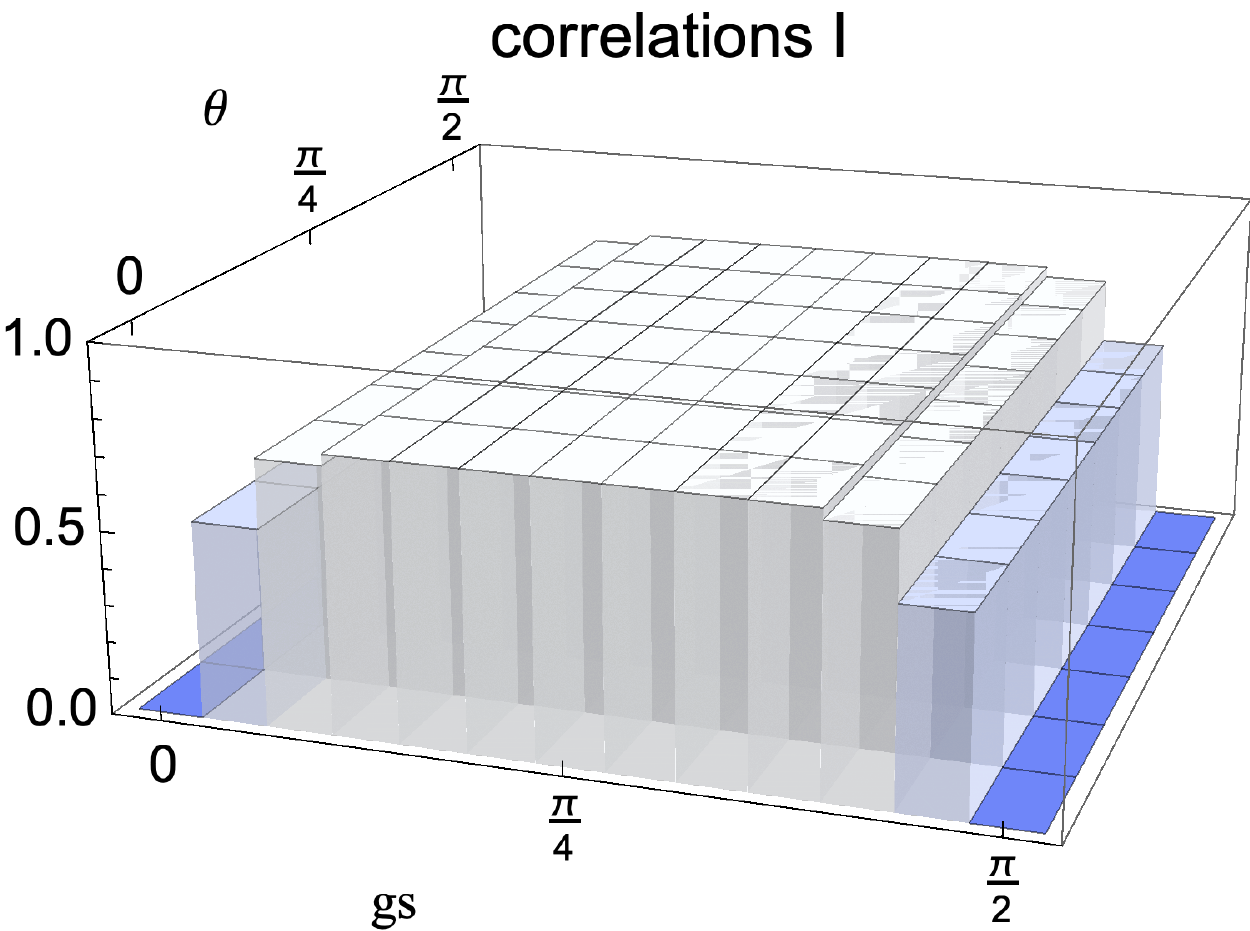}~~\includegraphics[width=0.24\textwidth]{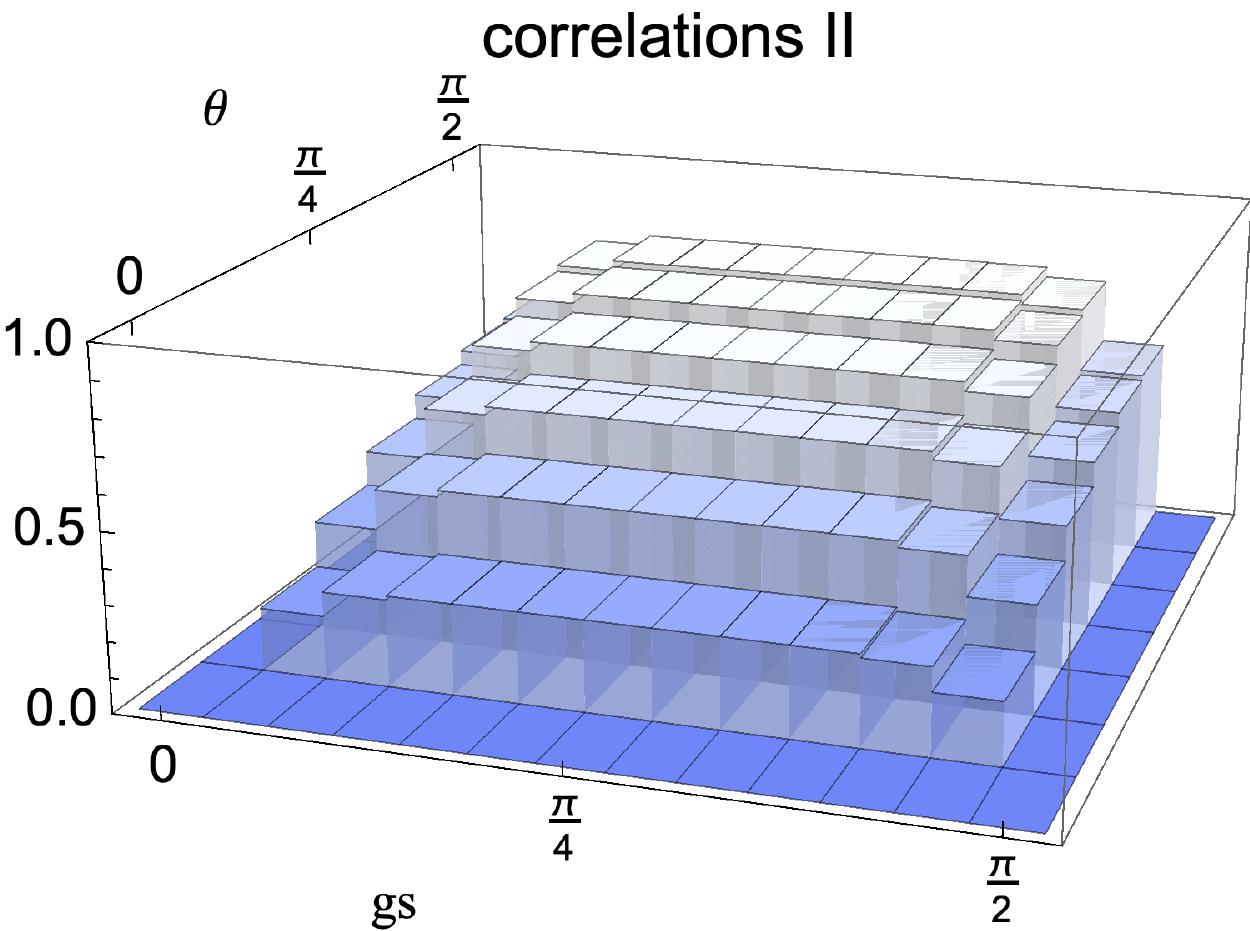}~~\includegraphics[width=0.24\textwidth]{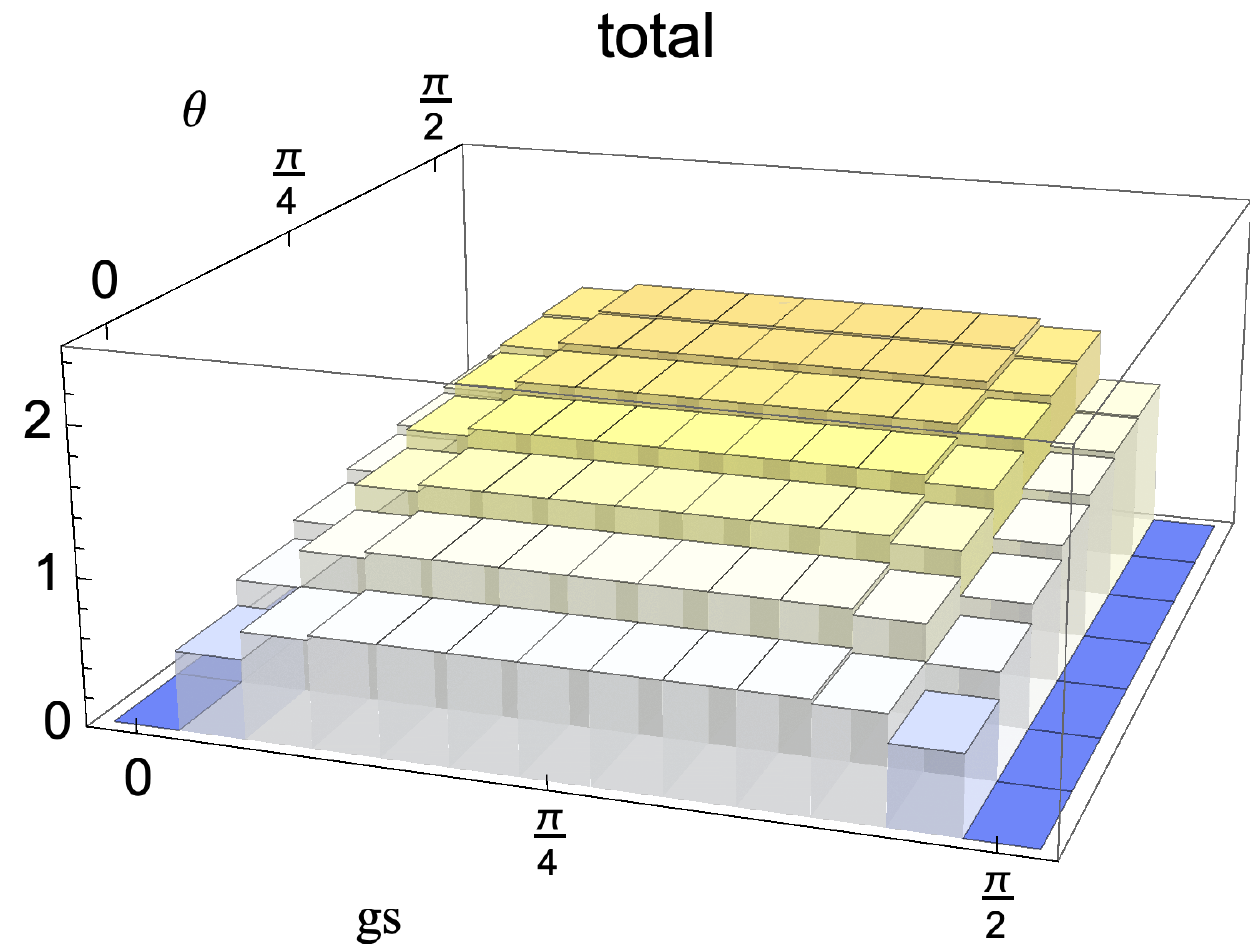}
\rule{\textwidth}{.1em}
\caption{Plot of the different contributions at the r.h.s. of
  Eq.~(\ref{eq:ineq}), together with their sum, all quantified via the
  QJSD$^{\sfrac 12}$.
They are considered as a function of running
time $gs$ and initial pair of states fixed by the angle $\theta$.
  The first row
  corresponds, as indicated, to the model determined by $c=\sfrac 12$,
  the second to $c=0$. For $c=\sfrac 12$ the environmental
  units have the maximum amount of coherence, while for $c=0$ they
  start in a maximally mixed state and the reduced environmental state
remains unchanged, so that one of the contributions is equal to zero.}
\label{fig:time}
\end{figure*}

\subsection{Model dependence of bounds on distinguishability
revivals}\label{sec:model-dep}

We first analyze the behavior in time of the bounds, exploring their
dependence on the considered model. In particular, we will investigate the models
arising for the choices $c = 0$ and $c=\sfrac 12$. We recall that the
non-Markovianity only depends on the behavior of the reduced state of the
system, so that it is exactly the same for all values of $c$. 
The revivals of
distinguishability for the whole class of models, expressed by means of the
QJSD$^{\sfrac 12}$, namely the l.h.s. of Eq.~(\ref{eq:ineq}), are shown in
Fig.~\ref{fig:lhs_time} as a function of the rescaled time and the choice of
initial system states. As expected, the highest revivals take place for
orthogonal initial states, corresponding to $\theta = \pi / 2$. The
periodicity of the dynamics, due to the uniform coupling, is also apparent. In
Fig.~\ref{fig:time} we show the behavior of the
contributions at the r.h.s. of the bound, which provide information on degrees
of freedom external with respect to the system, so that they are indeed model
dependent. We plot the contributions due to established correlations, starting
from the initial conditions $\rho^1_S (0)$ and $\rho^2_S (0)$, respectively,
together with the distinguishability of the correspondingly evolved
environmental states $\rho^1_E (s)$ and $\rho^2_E (s)$, as well as the sum of
the three terms which determines the overall tightness of the bound, Eq.~\eqref{eq:ineq}. The first
row shows the result for the model corresponding to $c=\sfrac 12$, in which
Darwinism appears, the second for $c = 0$. We see that in the first
case the upper bound is significantly less tight.
The main reason is that for
$c = 0$ the environmental state does not evolve, so that one of the
contributions is always zero, while in the other model changes of the
environmental dynamics take place for all choices of the $\theta$ parameter
different from $\pi / 2$, corresponding as discussed above to $\rho^2_S (0) =
| - \rangle \langle - |$. The correlations between the second reduced system
state and its environment, the only $\theta$ dependent ones, are
strongly affected by the parameter fixing this second initial system state, however in the
opposite manner with larger $\theta$ leading to more pronounced the correlations. As
a result the upper bound is only weakly $\theta$ dependent. Interestingly, the
maximum of the upper bound as a function of $\theta$ does not correspond to the
maximum of the bounded quantity, namely the l.h.s. of \ Eq.~(\ref{eq:ineq}),
shown in Fig.~\ref{fig:lhs_time} . In all cases, the dominant contribution is
given by the established correlations.

\subsection{Fraction dependence of bounds on distinguishability
revivals}\label{sec:frac-dep}

\begin{figure*}
  \begin{flushleft}
    $\mathbf{c=\sfrac 12}$
  \end{flushleft}
 \includegraphics[width=0.24\textwidth]{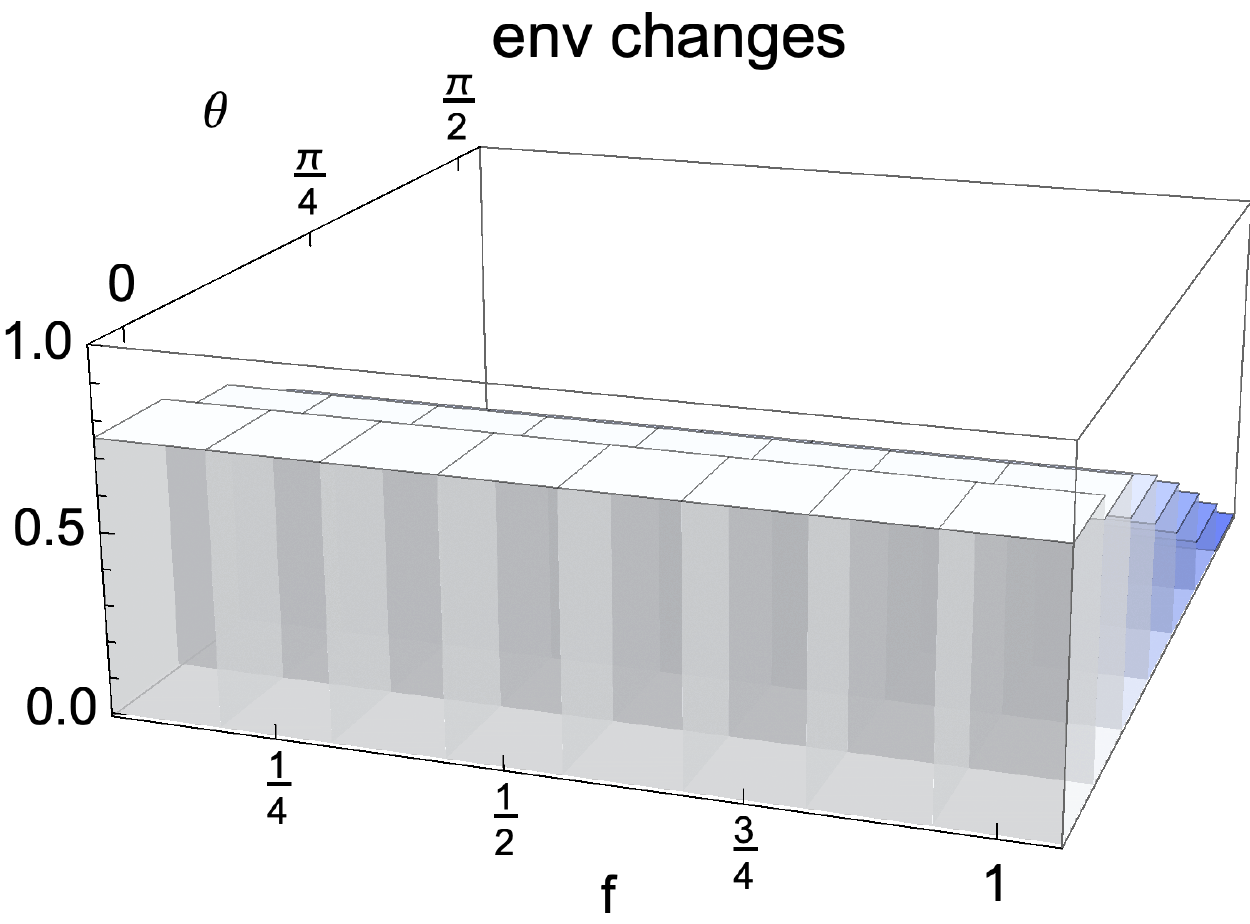}~~\includegraphics[width=0.24\textwidth]{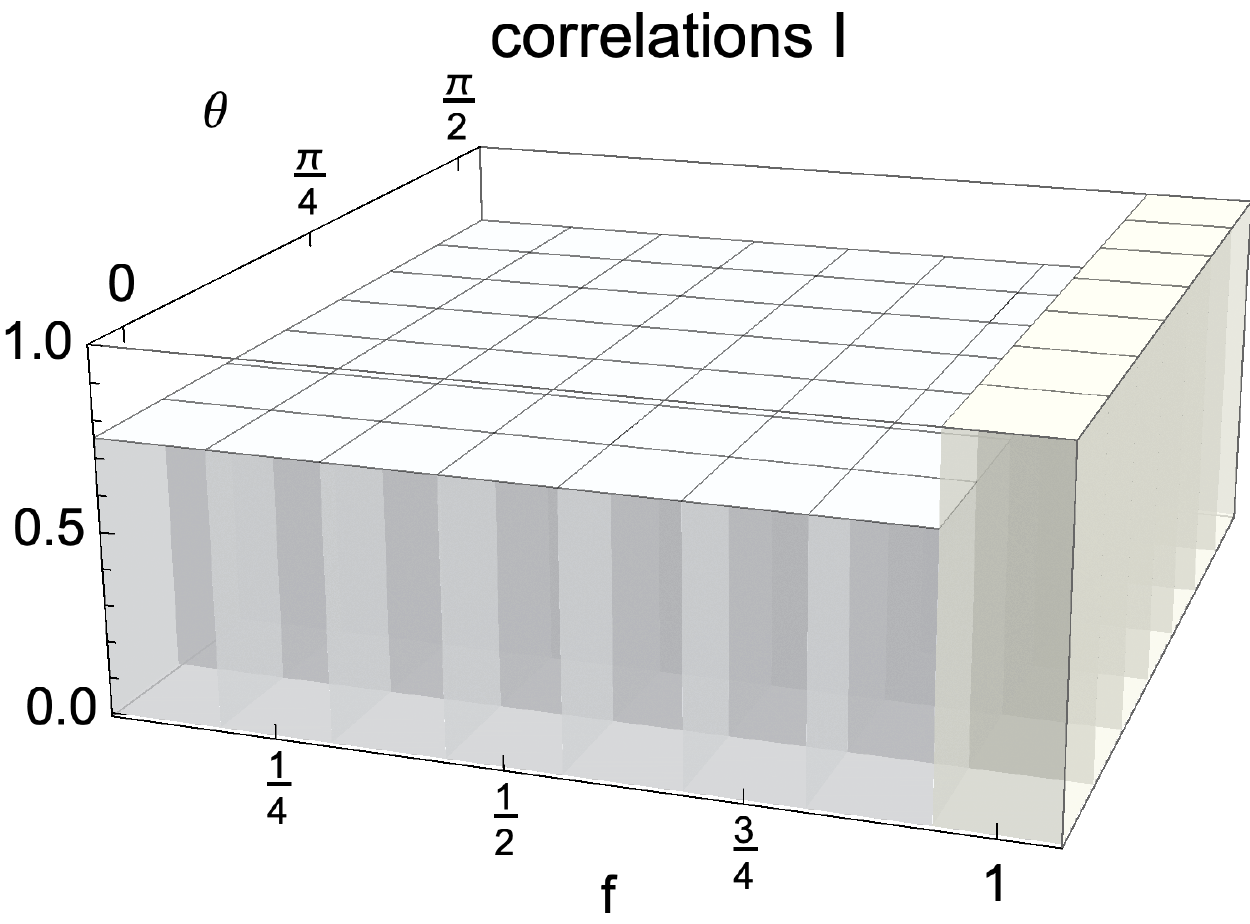}~~\includegraphics[width=0.24\textwidth]{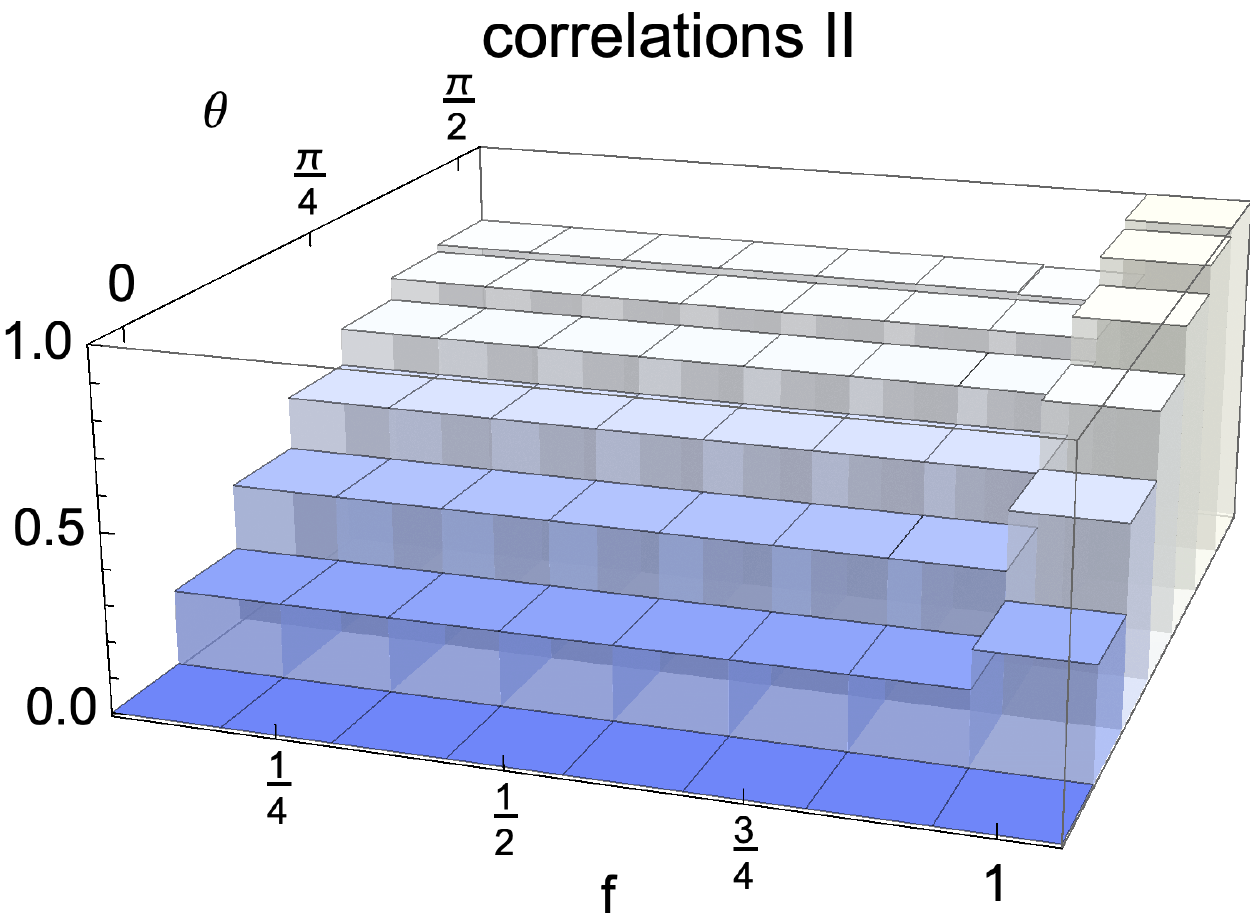}~~\includegraphics[width=0.24\textwidth]{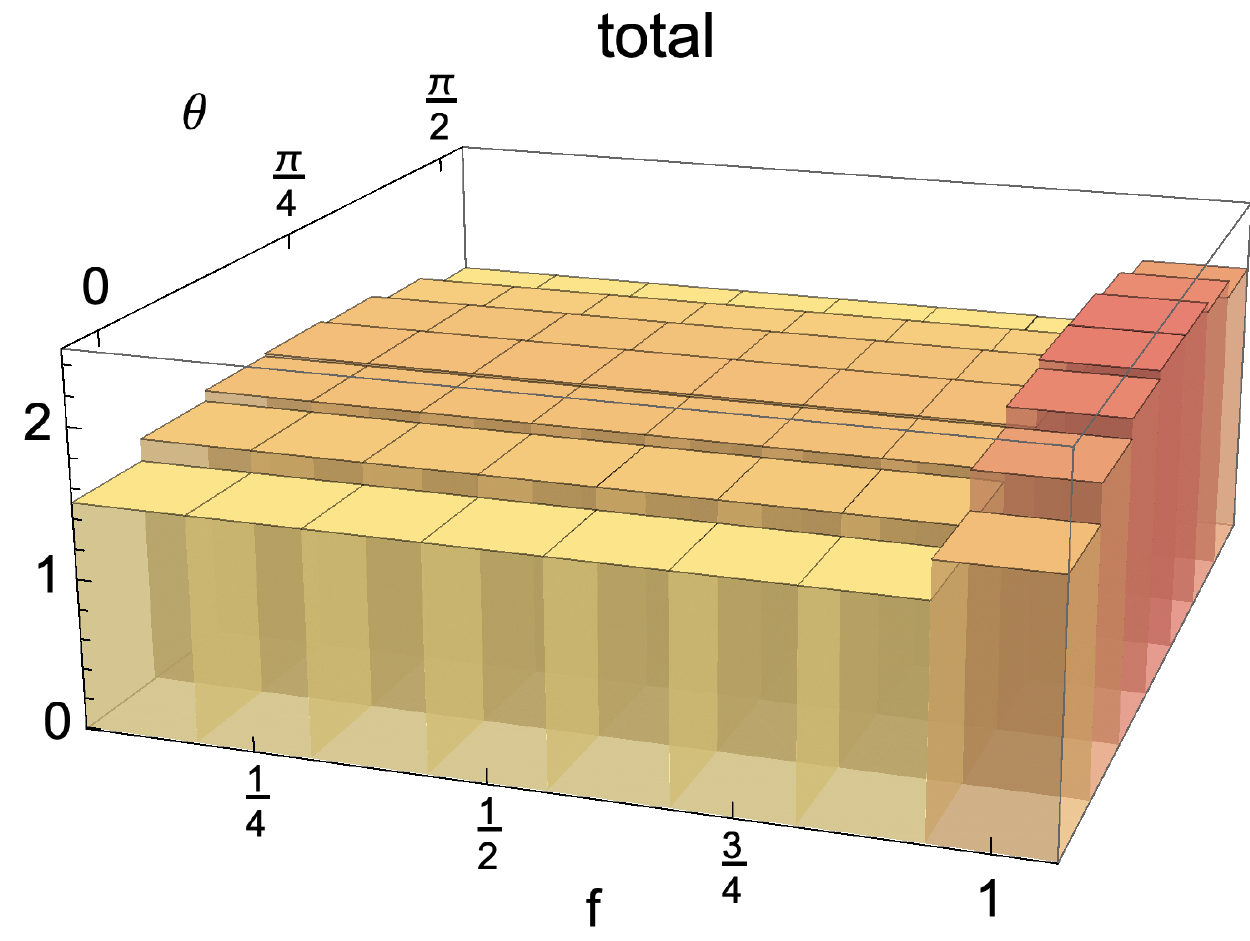}
\rule{\textwidth}{.1em}
 \begin{flushleft}
    $\mathbf{c=\sfrac 13}$
  \end{flushleft}
 \includegraphics[width=0.24\textwidth]{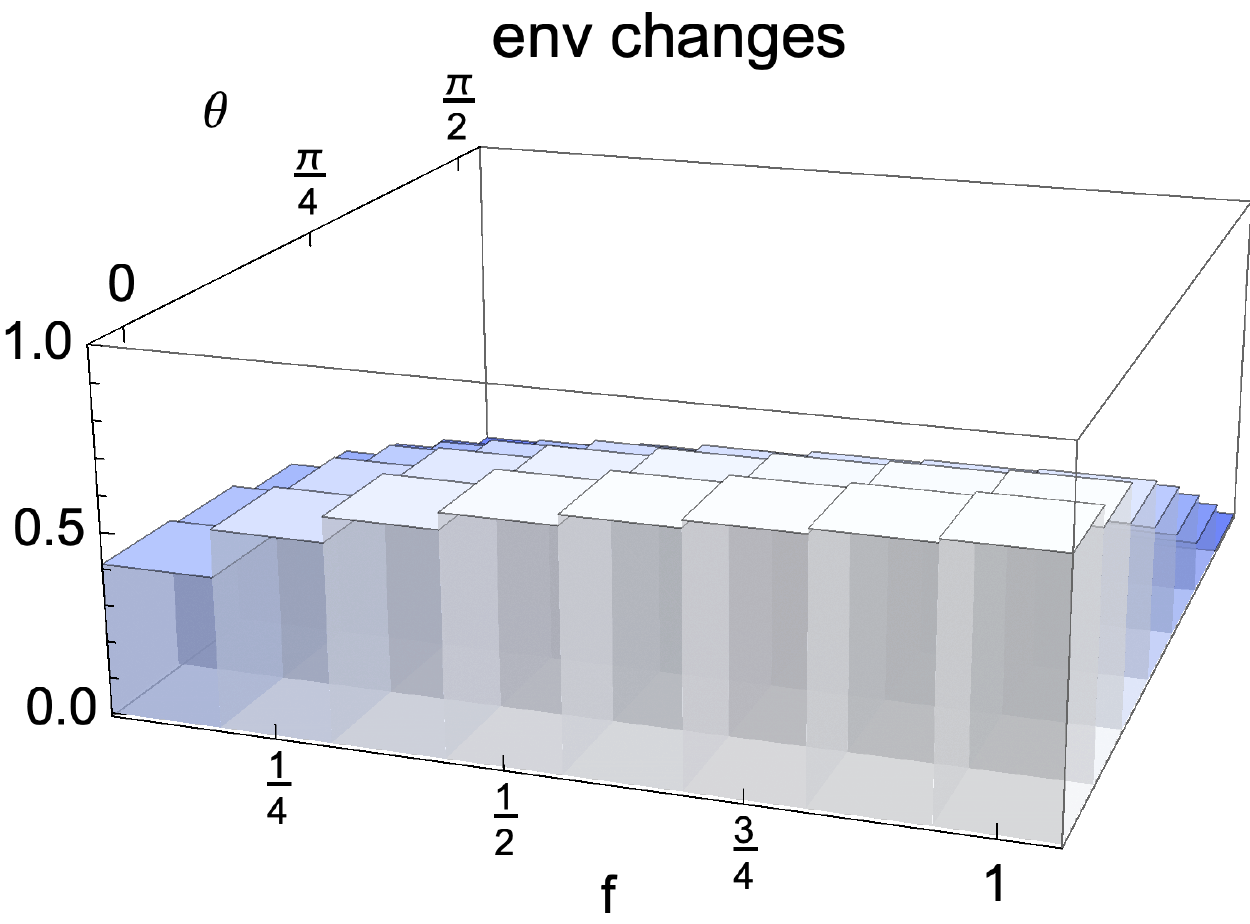}~~\includegraphics[width=0.24\textwidth]{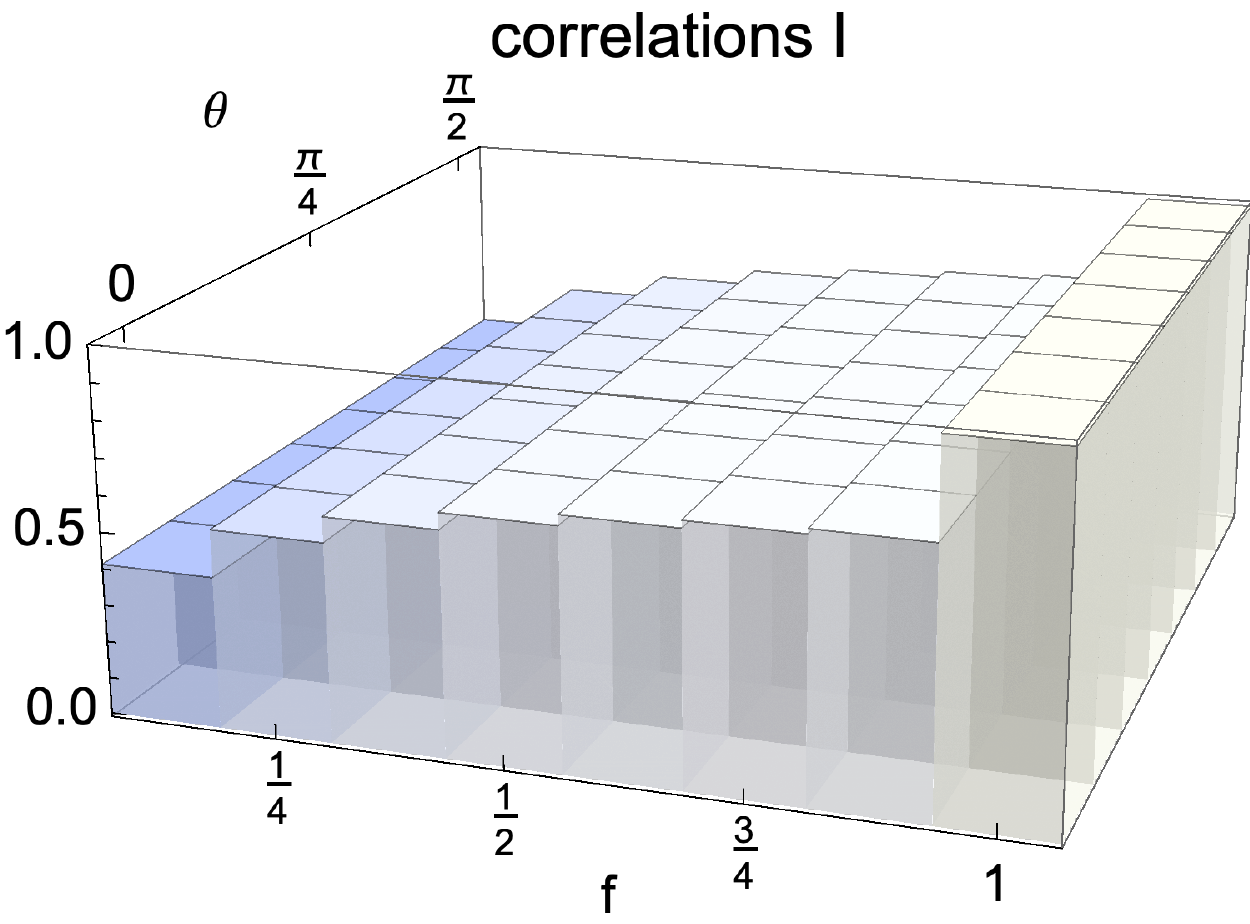}~~\includegraphics[width=0.24\textwidth]{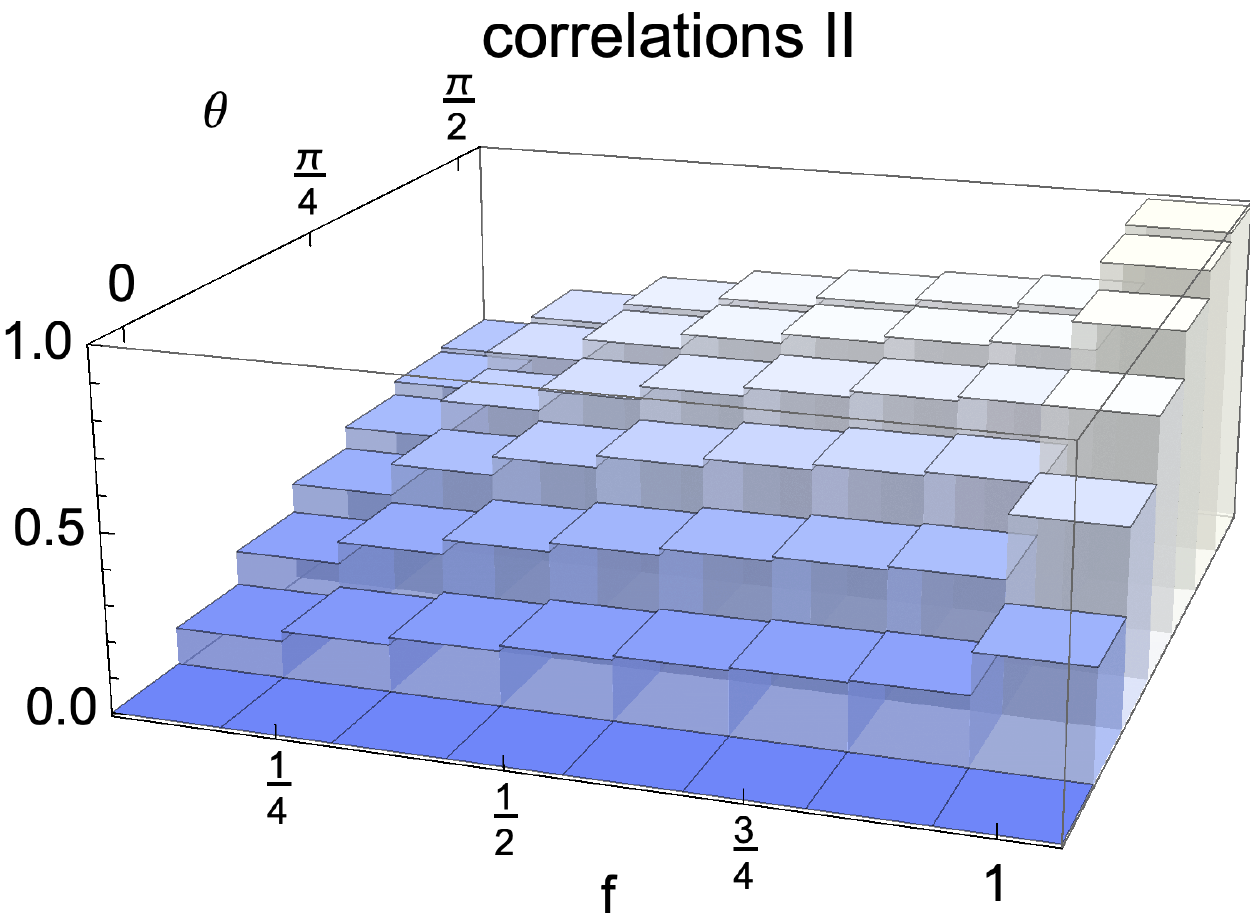}~~\includegraphics[width=0.24\textwidth]{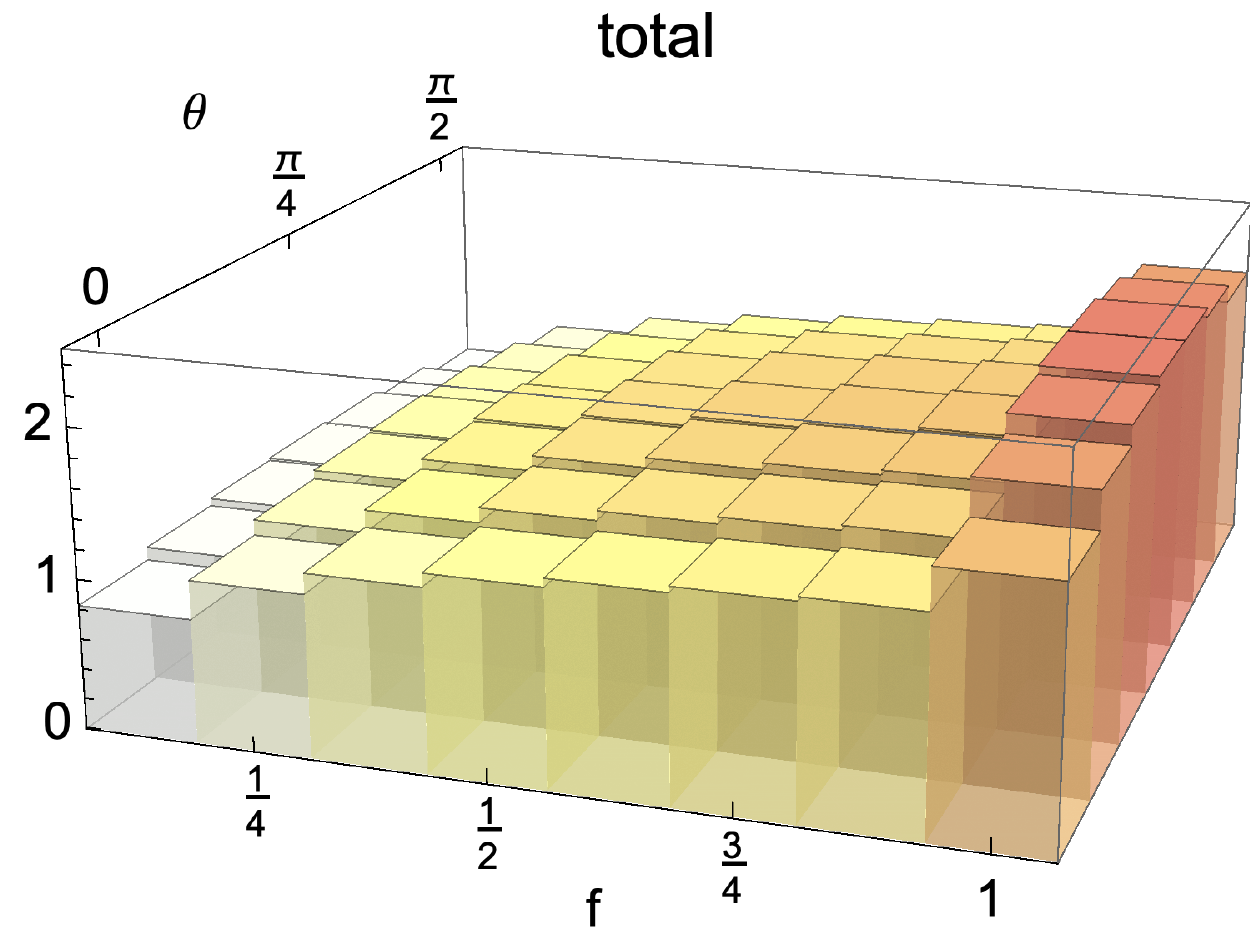}
\rule{\textwidth}{.1em}
 \begin{flushleft}
    $\mathbf{c=0}$
  \end{flushleft}
 \includegraphics[width=0.24\textwidth]{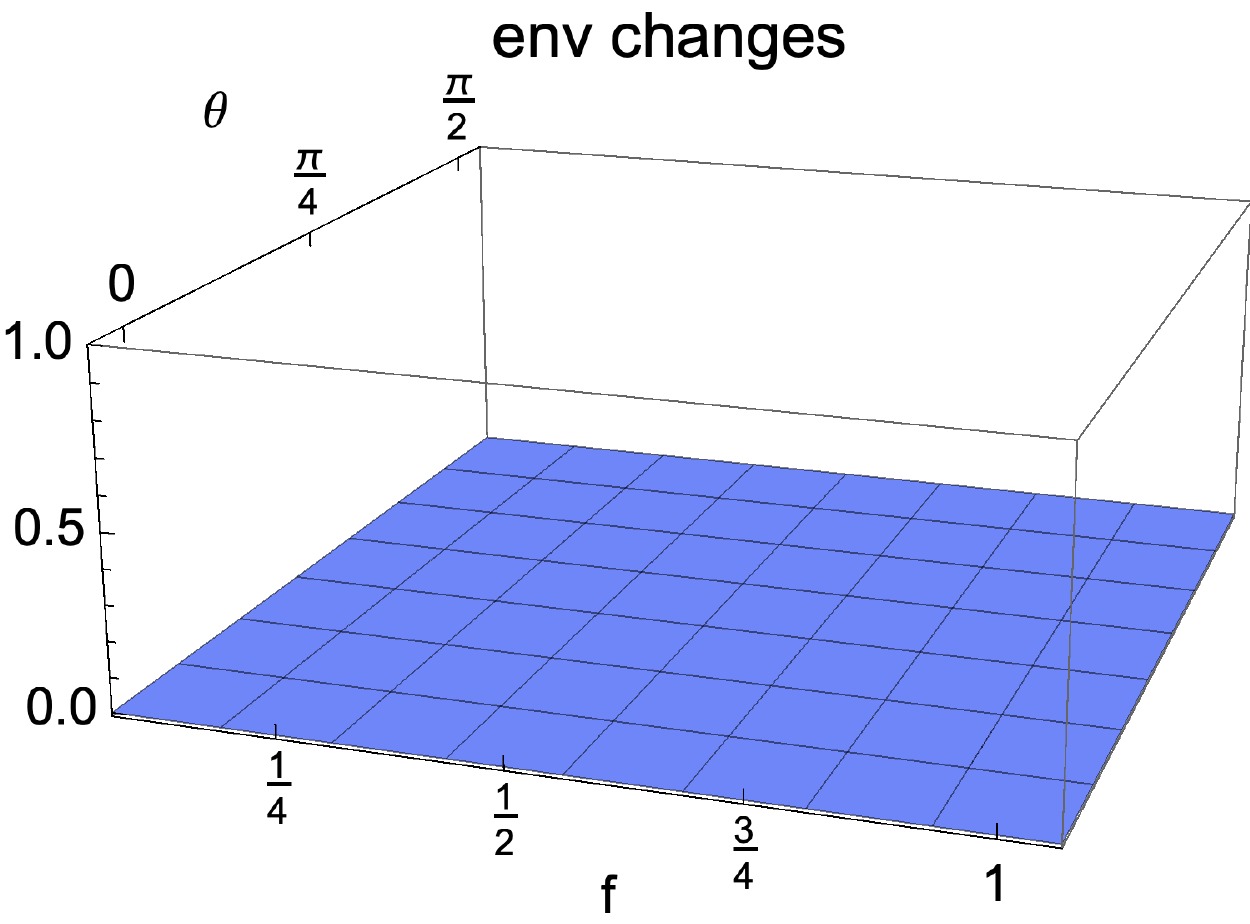}~~\includegraphics[width=0.24\textwidth]{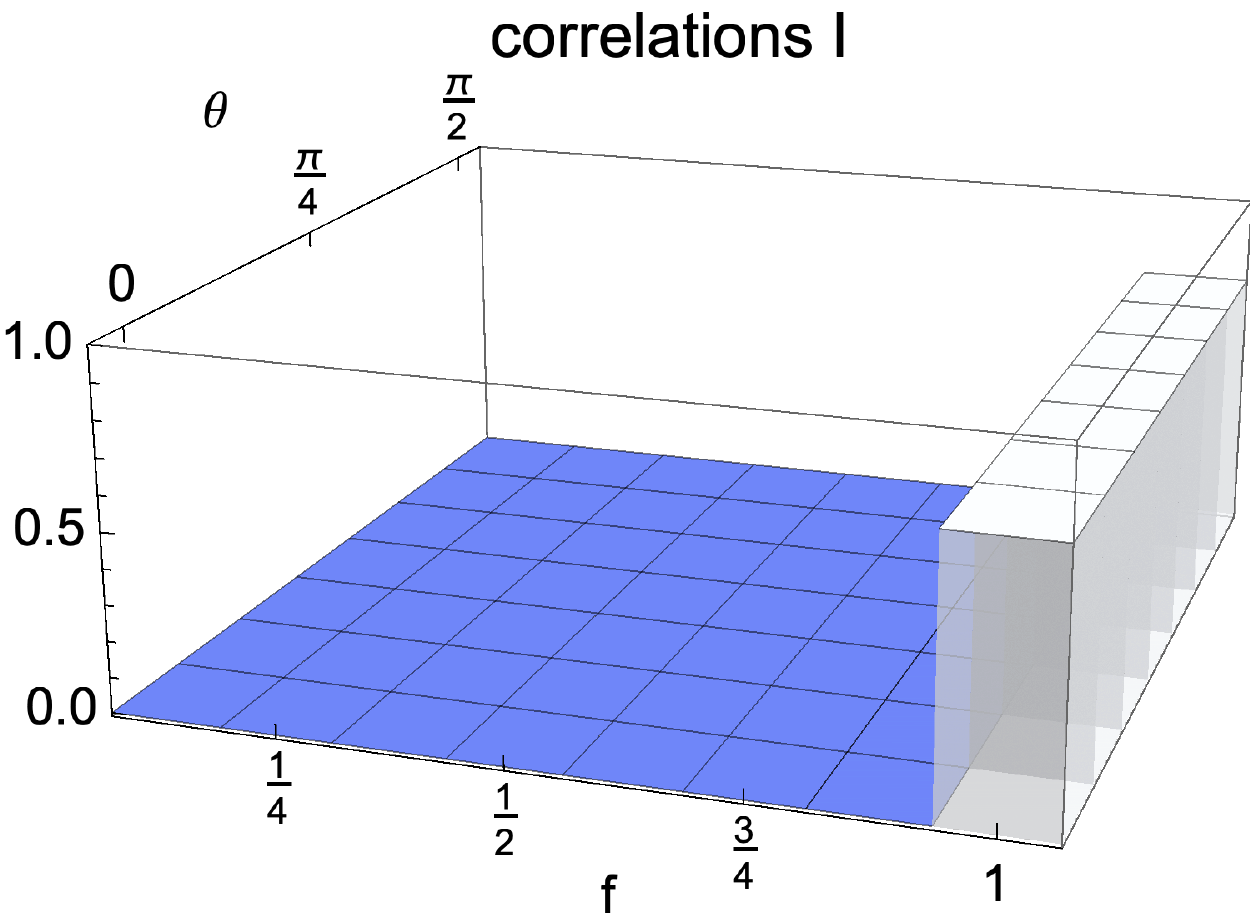}~~\includegraphics[width=0.24\textwidth]{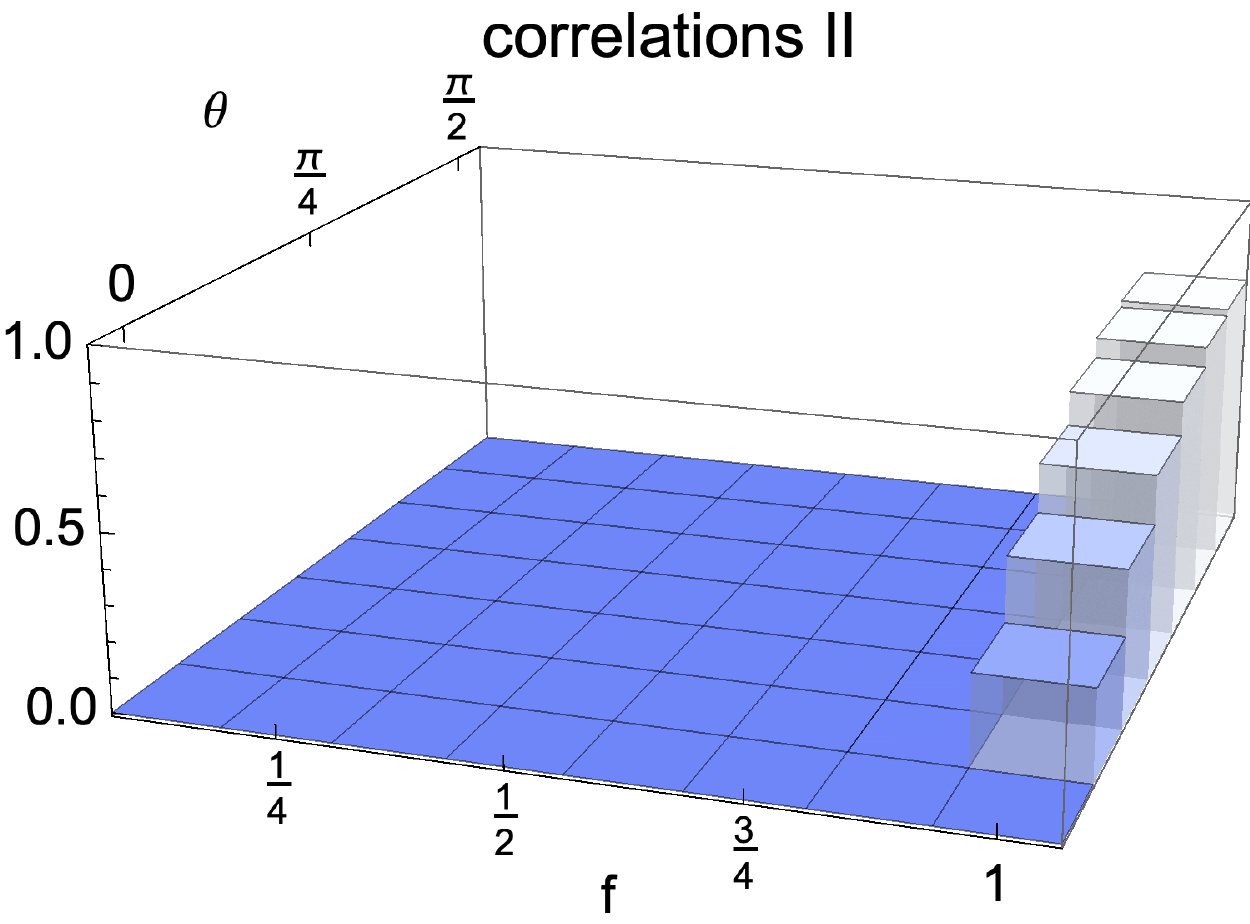}~~\includegraphics[width=0.24\textwidth]{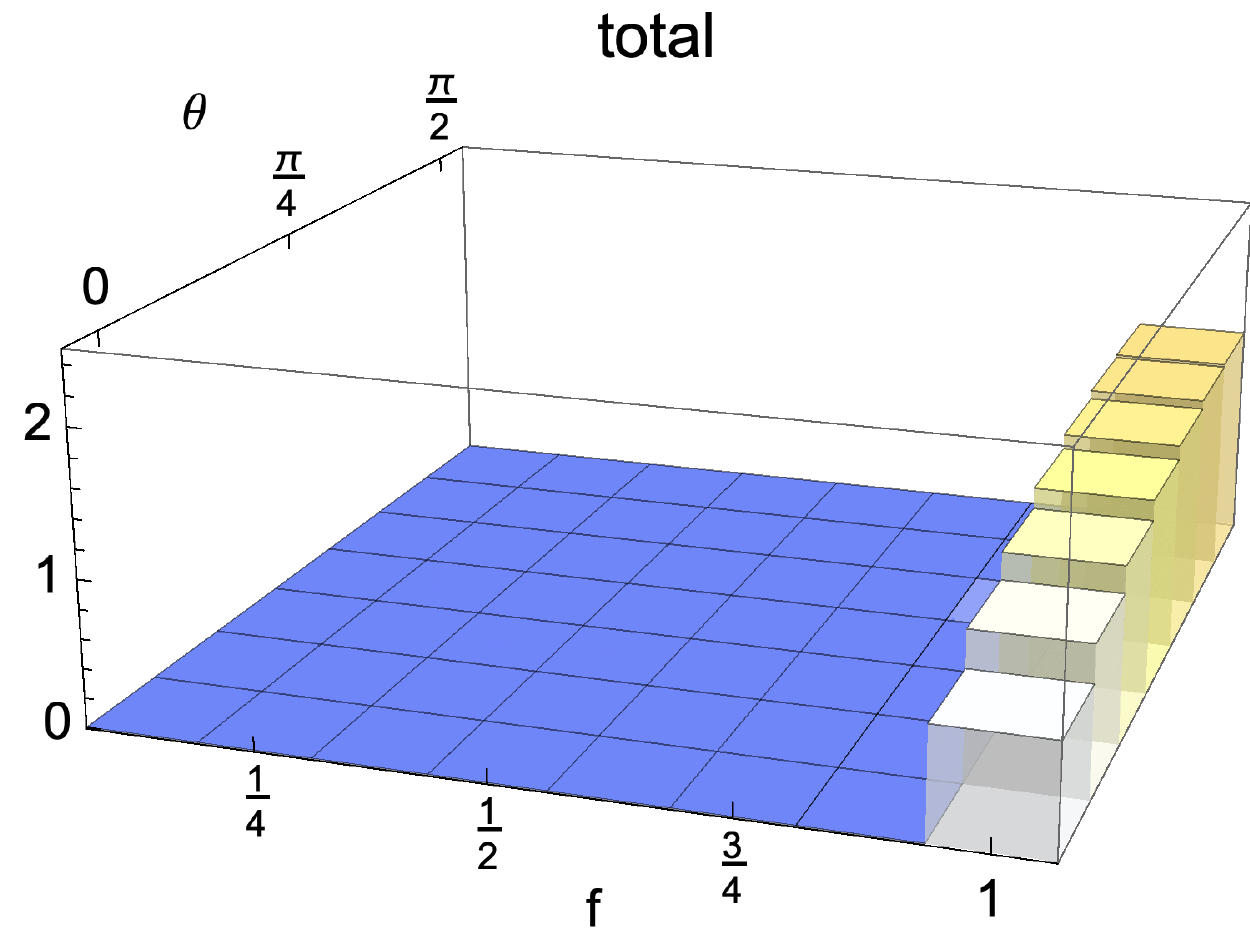}
\rule{\textwidth}{.1em}
\caption{Plot of the different contributions at the r.h.s. of
  Eq.~(\ref{eq:ineq}), together with their sum, evaluated for the case
  in which the total
  state is replaced by a marginal obtained by tracing out some
  environmental qubits, so that only a fraction ${\mathsf{f}}$ is
  considered. Also in this case all quantities are expressed via the
  QJSD$^{\sfrac 12}$.
They are considered as a function
of fraction ${\mathsf{f}}$
and initial pair of states determined by the angle $\theta$ for a fixed
time $g{s}=\pi/4$ .
  The first row
  corresponds, as indicated, to the model determined by $c=\sfrac 12$,
  the second to $c=\sfrac 13$  and the third to $c=0$. For $c=\sfrac
  12$ plateaus as a function of ${\mathsf{f}}$ are clearly observed, replaced for
  $c=\sfrac 13$ by a weak dependence. For $c=0$ a non zero valued is
  only obtained when including the whole environment, since tracing
  over any environmental units leads to a factorized state. 
  }
\label{fig:fraction}
\end{figure*}

\begin{figure}
\begin{center}
    \includegraphics[width=0.4\columnwidth]{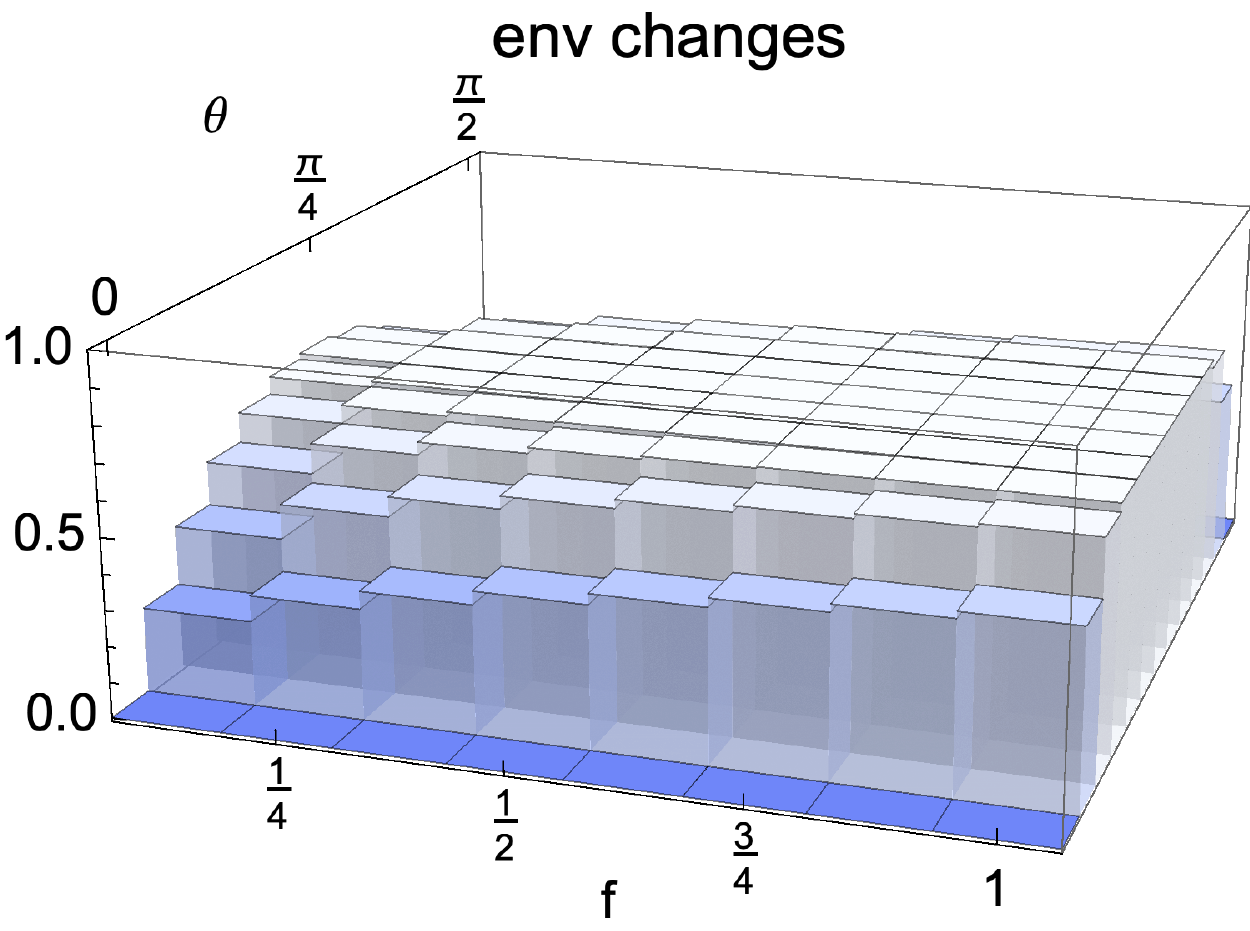}~\hspace{2em}~\includegraphics[width=0.4\columnwidth]{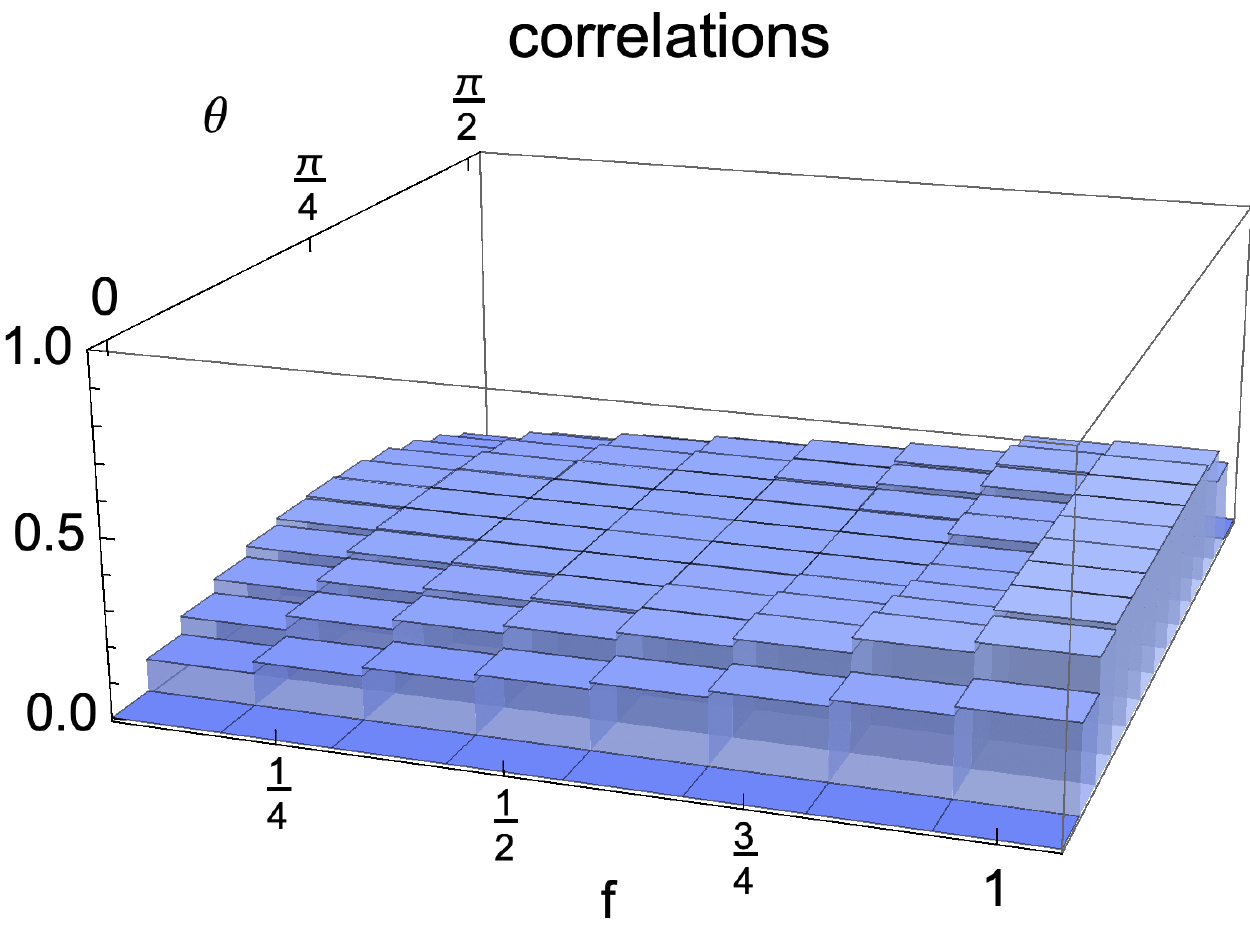}
\end{center}
    \caption{Behavior of environmental changes and correlations for
      the model with $c=\sfrac 12$ plotted as a function of both
      fraction ${\mathsf{f}}$ and time $gs$. It immediately appears
      that a plateau
      as a function of ${\mathsf{f}}$ only exactly takes place for the time
      $gs=\pi/4$, corresponding to full decoherence of the system.}
\label{fig:time_fraction}
\end{figure}

In order to understand the role of the spreading of information for the
description of non-Markovianity in the different models, we study the behavior
of the quantities at the r.h.s. of Eq.~(\ref{eq:ineq}) when replacing the
environment with a smaller one, given by a fraction of the original set of
degrees of freedom. To this aim we fix a reference time taken to be $gs = \pi
/ 4$, corresponding to full decoherence of the reduced system, when quantum
Darwinism is typically observed. Since the partial trace
is a completely positive trace preserving map, each contribution gets smaller
due to contractivity of the QJSD$^{\sfrac 12}$ under such maps, a feature
shared by all distinguishability quantifiers considered for the description of
memory effects. The inequality in Eq.~(\ref{eq:ineq}) is therefore no longer
required to hold true, since we are lowering the r.h.s. without affecting the
l.h.s. The model corresponding to $c = 0$, see last row of Fig.~\ref{fig:fraction}, is very special in this respect, as no information
whatsoever is stored in any fraction of the environment smaller than the total
environment, so that the bound is immediately violated. For all choices of
initial reduced states correlations are built solely with the total
environmental state, while by tracing out any number of environmental qubits
one gets a factorized state. Additionally, the maximally mixed environmental
state is invariant during the evolution for all choices of reduced initial
state, a property which is obviously preserved by taking into account only
some fraction of environmental degrees of freedom. On the other hand, in the
model obtained for $c=\sfrac 12$, so that the initial environmental states are
pure, as shown in the first row of Fig. \ref{fig:fraction} the difference in
environmental states does not depend on the fraction of the environment we are
taking into account.
To see the origin of this
behavior we come back to Eq.~(\ref{eq:c0t}) evaluated at time $gs = \pi / 4$
for $c=\sfrac 12$, which upon taking the partial trace with respect to system
and a fraction $\mathsf{f}$ of the environment leads to the state
\begin{align}
  \rho_{E_{{\mathsf{f} N}}} (\pi / (4g)) & =  \rho_{11} (0) \frac{1}{2^{{\mathsf{f} N}}} \bigotimes_{k =
  1}^{{\mathsf{f} N}} \begin{pmatrix}
    1 & - i\\
    i & 1
  \end{pmatrix} \nonumber \\ &+ \rho_{00} (0) \frac{1}{2^{{\mathsf{f} N}}} \bigotimes_{k = 1}^{{\mathsf{f} N}}
  \begin{pmatrix}
    1 & i\\
    - i & 1
  \end{pmatrix} \label{eq:rhoef},
\end{align}
whose only non zero eigenvalues are $\rho_{11} (0)$ and $\rho_{00} (0)$, so
that the difference in environmental states is not influenced by the number of
environmental units taken into account. We stress that this is only true for
$gs = \pi / 4$, the point in time most relevant for the study of quantum
Darwinism. This can be seen considering both the dependence on time and
fraction as in Fig. \ref{fig:time_fraction}, where both environmental changes
and correlations are considered. Remarkably, for this particular model the
occurrence of a plateau as a function of the environmental fraction is not
only true for the changes in the environment, but also in the correlations
and, consequently, in the sum of these three quantities providing the overall
bound.
{In other words, the information exchange relevant for the onset of non-Markovianity only involves a small portion of the environment, so that the bound holds for any considered fraction.}
The appearance of these plateaus makes the dynamics indeed compatible
with quantum Darwinism, even though it only provides a sufficient condition
for the redundant spreading of information.

\begin{figure}
\begin{center}
  \includegraphics[width=0.45\columnwidth]{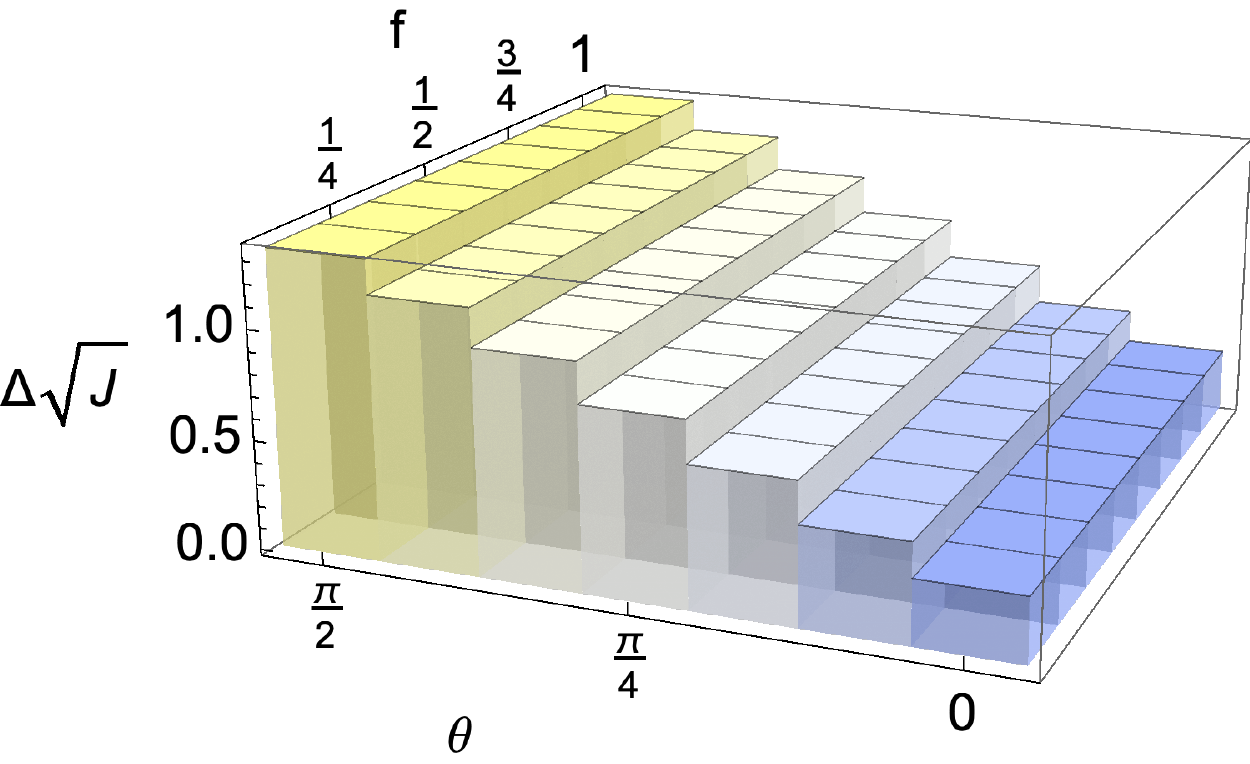}\hspace{1em} 
    \includegraphics[width=0.45\columnwidth]{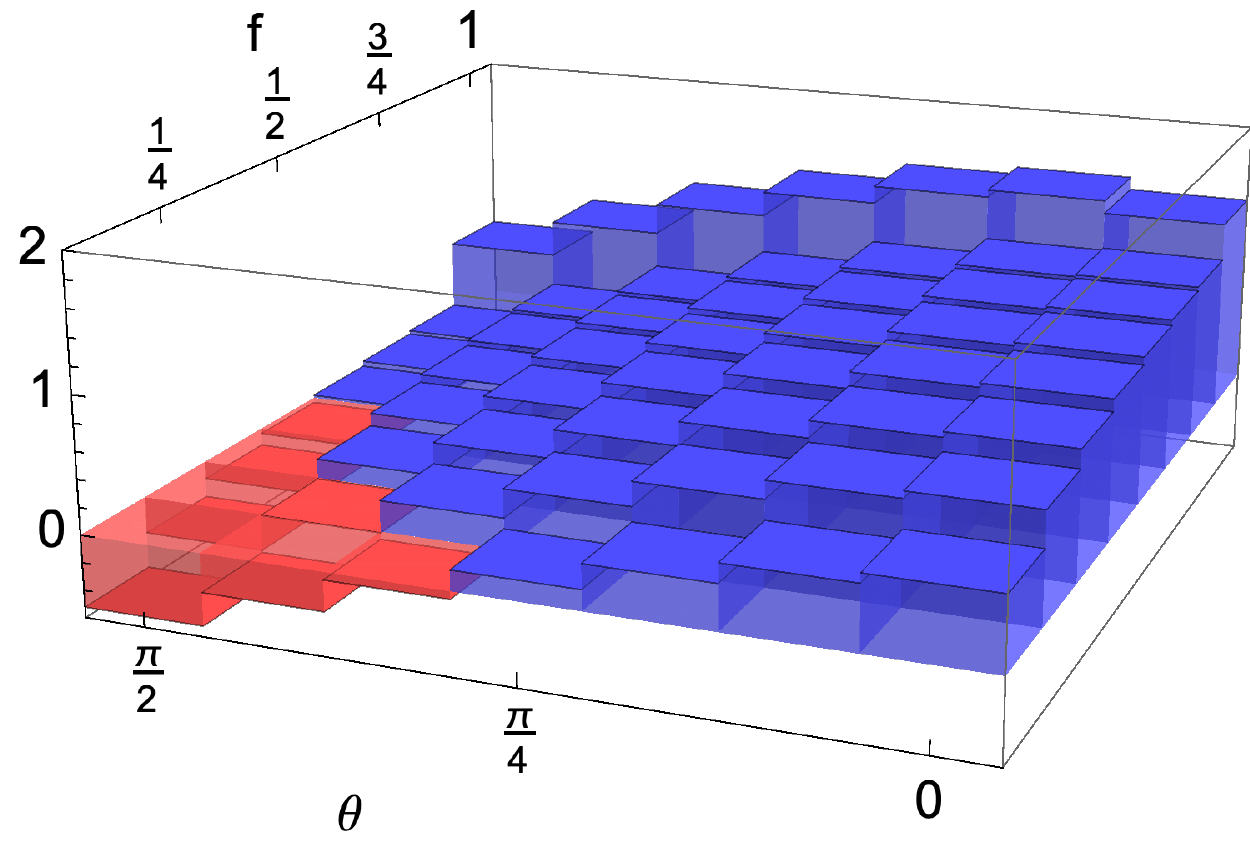} 
\end{center}
 \caption{{Left: Plot of the l.h.s. of Eq.~(\ref{eq:ineq})
               showing the revivals of the QJSD$^{\sfrac 12}$ as a
               function of 
               choice of initial system states for the fixed times $gt=\pi/2$ and 
               $gs=\pi/4$, the latter corresponding to maximal decoherence of the
               system. The quantity inherently does not depend on ${\mathsf{f}}$. Right: The difference of the l.h.s. and sum of quantities on the r.h.s. of the inequality given by Eq.~(\ref{eq:ineq}) when taking into account only a fraction ${\mathsf{f}}$ of the environment, for $c=\sfrac 13$ (see Fig. \ref{fig:fraction}, last figure of the second row). For the values of environmental fraction $\mathsf{f}$ and angle $\theta$ (which determines the pair of initial system states)
               for which the difference is negative (red) the corresponding sum of environmental changes and correlations is no longer an upper bound for the revivals in the reduced dynamics.} 
               }
\label{fig:lhs_fraction}
\end{figure}

The occurrence of a very weak dependence with respect to the stepwise
inclusions of environmental degrees of freedom is not new, it was indeed
already observed in a collisional framework {\cite{Campbell2019b}}. It reflects
the fact that given the size of the system, the correlation with a small
portion of the environment is already sufficient to store the information
necessary to lead to a revival in distinguishability of the system states. In the
present framework for $c=\sfrac 12$ we are faced with a true plateau, which
reflects the pure Darwinistic behavior exhibited by this model. To better
clarify this behavior we have also plotted the same
quantities for an intermediate choice of the mixing parameter $c=\sfrac 13$, see
middle row of Fig. \ref{fig:fraction}. In this case, corresponding to a model
in which Darwinism is partially washed out {\cite{Zwolak2009a}}, a weak
dependence on the size of the fraction can be observed, so that a larger part
of the environment is necessary to recover the relevant information. For this
model the failure of the upper bound upon tracing out part of the environment
can be observed for a small enough fraction and large values of the parameter
$\theta$,  {see Fig. \ref{fig:lhs_fraction}. While the amount of information flowing back to the open system does not depend on the
fraction of the environment taken into account, the capability to trace it back to the established 
correlations between the system and a portion of the environment, as well as to the changes of the latter,
requires now that a portion large enough is considered.}

\section{Conclusions} \label{sec:conc}
We have investigated the subtle role that system-environment correlations play in the characterisation of a given dynamics. Through a paradigmatic class of dephasing models, which are particularly relevant in exploring the quantum Darwinism framework, we employed tools from the study of non-Markovianity to critically assess the role that these correlations play, revealing that while only a small amount of such correlations are needed for the on-set of non-Markovianity, establishing the conditions for classical objectivity necessitates significantly more. {Our results indicate that for most microscopical realisations of the reduced dynamics} one can fully capture the non-Markovian characteristics of a given evolution with access to only a small subset of the environmental degrees of freedom; while also revealing that whether the conditions for classical objectivity are satisfied or not is crucially dependent on the precise details of the microscopic model in question, rather than its non-Markovian nature.








\begin{acknowledgments}
N.M. was funded by the Alexander von Humboldt Foundation in form of a Feodor-Lynen Fellowship and project ApresSF, supported by the National Science Centre under the QuantERA programme, which has received funding from the European Union's Horizon 2020 research and innovation programme. S.C. gratefully acknowledge the Science Foundation Ireland Starting Investigator Research Grant ``SpeedDemon" (No. 18/SIRG/5508) for financial support. A.S. and B.V. acknowledge support from UniMi, via Transition Grant H2020 and PSR-2 2020.
\end{acknowledgments}

\end{document}